\documentclass[10pt]{article}

\usepackage{times,newtxmath}
\usepackage{graphicx}
\usepackage{tabu,booktabs}
\usepackage{enumitem}
\usepackage{mathtools}
\usepackage{algorithm,algpseudocode}
\usepackage{subfig}
\usepackage{url}

\usepackage{pbltx}

\newcommand{\Twt}{x}
\newcommand{\Ent}{e}
\newcommand{\Ents}{\mathcal{E}}
\newcommand{\Tpc}{t}
\newcommand{\Tpcs}{\mathcal{T}}
\newcommand{\pEntTpc}{p_{\Ent,\Tpc}}
\newcommand{\qEntTpc}{q_{\Ent,\Tpc}}
\newcommand{\nEntTpc}{n_{\Ent,\Tpc}}
\newcommand{\sEntTpc}{s_{\Ent,\Tpc}}

\newcommand{\Int}{i^u}
\newcommand{\IntTpc}{\Int_\Tpc}
\newcommand{\Exp}{v}
\newcommand{\Exps}{\mathcal{V}}

\newcommand{\TpcExp}{\theta^\Tpc}
\newcommand{\Listed}{\lambda}
\newcommand{\Followers}{\phi}

\newcommand{\ExpFols}{\Exps^u}
\newcommand{\wExpTpc}{w^{\Exp}_{\Tpc}}
\newcommand{\TpcInTpcsExp}{\Tpc \in \Tpcs^\Exp}
\newcommand{\ExpInExpFols}{\Exp \in \ExpFols}

\DeclareMathOperator{\OddsRatio}{OddsRatio}
\DeclareMathOperator{\Multinomial}{Multinomial}
\DeclareMathOperator{\fRank}{f_{\text{rank}}}
\DeclareMathOperator*{\argmax}{arg\,max}

\newenvironment{NewText}{}{}

\newcommand{\todoNG}[1]{}
\newcommand{\citeN}[1]{\cite{#1}}

\begin{document}

\title{What You Like:
  Generating Explainable\\
  Topical Recommendations for Twitter\\
  Using Social Annotations}

\author{Parantapa Bhattacharya, Saptarshi Ghosh, Muhammad Bilal Zafar,\\
Soumya K. Ghosh, and Niloy Ganguly}

\maketitle

\begin{abstract}
With over 500 million tweets posted per day, in Twitter,
it is difficult for Twitter users
to discover interesting content
from the deluge of uninteresting posts.
In this work,
we present a novel, explainable, topical recommendation system,
that utilizes social annotations,
to help Twitter users discover tweets,
on topics of their interest.
A major challenge
in using traditional rating dependent recommendation systems,
like collaborative filtering and content based systems,
in high volume social networks is that,
due to attention scarcity most items do not get any ratings.
Additionally, the fact that most Twitter users are passive consumers,
with 44\% users never tweeting,
makes it very difficult to use user ratings
for generating recommendations.
Further, a key challenge
in developing recommendation systems is that
in many cases users reject relevant recommendations
if they are totally unfamiliar with the recommended item.
Providing a suitable explanation,
for why the item is recommended,
significantly improves the acceptability of recommendation.
By virtue of being a topical recommendation system
our method is able to present
simple topical explanations
for the generated recommendations.
Comparisons with state-of-the-art
matrix factorization based collaborative filtering,
content based and social recommendations
demonstrate the efficacy of the proposed approach.
\footnote{%
This study was conducted during the year 2016
when Parantapa Bhattacharya was a student at 
Indian Institute of Technology Kharagpur.

This study was conducted respecting the guidelines
set by our institute's ethics board
and with their knowledge and permission.}
\end{abstract}

\section{Introduction}
\label{sec:intro}

As social media and user generated content
have gained popularity,
so has the problem of information overload.
In Twitter, over 500 million tweets
are posted per day~\cite{500M-tweets-daily}.
This makes it incredibly difficult for users
to discover new and interesting content
from within the deluge of uninteresting ones.
To help users navigate through massive amounts of content
and to direct their attention
towards items that would potentially be of their interest,
designers of online systems
have traditionally used recommender systems.
Recommender systems have been developed
to help users in numerous contexts,
including item purchase~\cite{schafer-ec99},
movie selection~\cite{miller-iui03},
and choosing music~\cite{yoshii-taslp08}.
A large volume of literature has also been published
to help users find better content
in online social media systems
\cite{chen-sigir12,pan-recsys13,forsati-tois14,jiang-tkde14,chaney-recsys15,jiang-tkde15}.

Most of the state-of-the-art recommender systems,
including those discussed above,
are powered by collaborative-filtering~\cite{shi-csur14}
or are content-based recommender systems~\cite{lops-book11};
and sometimes a combination of both.
To be successful,
collaborative-filtering and content based methods
depend on having rating or like/dislike information
associated with large number of users and items.
However, with posts being generated at such huge volumes,
attention scarcity does not allow
most content to receive any likes or retweets in Twitter.
We estimate that less than 13\% tweets receive any retweets
and less than 23\% tweets receive any likes or favorites
(Section~\ref{ssec:response-estimate}).
The problem is compounded by the fact that
44\% Twitter users are only passive consumers of content,
never posting any tweets themselves~\cite{Never-tweeters}.
The problem is similar for other high volume social media
systems such as Reddit~\cite{gilbert-cscw13}.
Hence, for such systems, the well-known cold start problem
is present not only for new users
but persists for a large fraction of old but passive users.
For content, however, the cold start problem is even worse,
as the cold start problem keeps renewing itself eternally,
due to huge content creation rates and small content lifetimes.

Another issue in developing recommender systems
for online social networks like Twitter is that
tweets that eventually become popular,
receiving likes or retweets,
take time to accumulate these interactions.
As one of the prime focus of microblogging platforms, like Twitter,
is to deliver information in realtime,
it is problematic to build recommender systems for these platforms
that depend on such delayed information as part of their core methodology.
Additionally, state-of-the-art recommender algorithms such as
matrix factorization based collaborative-filtering
are quite computationally intensive.
While this is acceptable for systems
where the item set is relatively stable over time,
it becomes inadequate when over 500 million items
are created during the period of a single day.

A common problem for developers of recommender systems is that,
their users are in many cases
hesitant to accept recommendations
that they are completely unfamiliar with,
even when they are relevant~\cite{herlocker-cscw00}.
An explanation for why the item was recommended
significantly increases the probability
of the user accepting the recommendation.
This problem becomes even more acute
when recommending tweets,
which due to their short lengths,
can be difficult to interpret without context.
What further compounds the problem of explainability is that
traditional collaborative-filtering and content based recommender systems
encode user interests using latent representations
which are difficult for humans to interpret.
Studies which have tried to explain recommendations
have generally disassociated
the process of generating recommendations
and the method of explaining them~\cite{herlocker-cscw00}.
However, this disassociation between
recommendation generation and recommendation explanation procedures,
can lead to users developing wrong intuitions
about the recommendation methodology,
leading to lower trust and dissatisfaction
when the intuition is violated~\cite{eslami-chi15}.

In this work, we try to work around the above difficulties,
by developing a novel personalized recommender system for Twitter,
that utilizes social annotations
to infer interests of Twitter users and ascertain topics of tweets.
Social annotations are pieces of information about a user,
that is mined from how others describe the given user.
For example, in this current work we mine
social annotations from Twitter Lists feature.
Using Twitter Lists, regular Twitter users can
create named groups
for organizing their social connections.
But in doing so they describe the members of such groups.
\citeN{sharma-wosn12} used social annotations in Twitter
to infer descriptive tags for a large number Twitter users.
In our previous work \cite{bhattacharya-recsys14},
we presented a preliminary work
trying to infer a user's relative interests in different topics.
This was done by simply comparing the number of experts
on the different topics
that the user was following.
Social annotations have also been used by other authors, including:
to establish connections between a user's follow relations~\cite{garcia-silva-eswc12},
and to analyze a user's political leanings~\cite{boutet-snam13}.

This work utilizes the intuitions presented in our earlier work
\cite{bhattacharya-recsys14}
and further improves on them
to create a novel topical recommendation system
which by virtue of its design is highly explainable.
At a high level, our method for generating personalized recommendations
can be conceptualized as a four-step process.
\emph{First}, using social annotations
we mine meaningful topics of expertise
for a large corpus of Twitter users (whom we refer to as topical experts).
\begin{NewText}
\emph{Second}, for every user for whom we generate recommendations,
we infer their interests in different topics
by modeling their followings of topical experts on different topics.
For this purpose we propose a variant of the Preferential Attachment
model~\cite{barabasi-science99},
that we call the Topical Preferential Attachment model,
and present an Expectation-Maximization based algorithm
that utilizes it to infer user interests.
\emph{Third}, we present a method for efficiently computing
relative relevance of large volumes of tweets
with respect to given topics.
For this purpose, we propose a tweet ranking methodology,
that we call the Topical Binary Independence Model,
which uses the content of tweets posted by topical experts on a given topic,
to estimate the relevance of a tweet with respect to the given topic.
\end{NewText}
\emph{Finally}, to generate recommendations for a user,
we present her with tweets that are most likely
to be about her topics of interest.

The above approach has a number of advantages
over traditional collaborative filtering and content based
recommender systems.
\begin{itemize}
\item \emph{First}, as the core methodology does not depend on user ratings,
it can be used for recommending to the large fraction of passive users.
Also, this makes it suitable for recommending items that may not be  popular
but are potentially interesting to different topical communities.
\item \emph{Second}, in social networking systems such as Twitter,
where huge volumes of content is being created constantly
and interest in created items fades quickly,
information mined from user-item relationships
deprecate quickly.
For example, a user may tweet about her political opinions
during a presidential election cycle,
however she may only have minor interest in politics
and may not be interested in politics most of the time.
Thus, the use of a higher level abstraction
of mining user-interest relations
has an added advantage of inferring more
stable relationships.
\item \emph{Finally}, not using ratings has the added advantage
that relevance of tweets to topics can be computed,
without having to wait for likes and retweets to accumulate.
Another, major advantage of the method is that,
because of its approach
generating explanations of the form
``this tweet is on topic $\Tpc$
and it has been recommended to you
because you seem to be interested in $\Tpc$''
becomes very straight forward.
Additionally, since such an explanation is easy to understand,
it is expected to increase user satisfaction.
\end{itemize}

However,  a number of challenges need to be tackled
to successfully implement and develop such a system.
\emph{First}, using social annotations
as one of the primary sources of information,
for building recommendation systems,
requires that a very large corpus of experts be identified,
with expertise on a diverse set of topics.
Size of the expert set bounds the set of users
for whom interests can be inferred,
while diversity of the expert set
limits the range of interests that can be inferred.
Further, having a sizable set of experts for different topics is essential
in trying identify robust signals
of what characterizes information on those topics.
\emph{Second}, while previous
works~\cite{ghosh-sigir12,bhattacharya-recsys14,zafar-cscw16}
have inferred tags of expertise from social annotations,
the inferred tags were very noisy.
For example the topic `Celebrity' was present in the corpus
through  a number of abbreviated and mispronounced tags
such as: celeb, famous, celbrity, etc.
Presence of such tags was not a major hindrance for
\cite{ghosh-sigir12} and \cite{zafar-cscw16},
as they were primarily search systems,
where matching user query (even if misspelt) to results was more important
than the cleanliness of the tags themselves.
However, as the present work is a recommendation system
and the topics identified are displayed to the users
as part of the explanation process,
it is important to have clean topics and not noisy tags.
\emph{Third}, \cite{bhattacharya-cscw14} had shown
that topics identified via social annotation have wide range or popularity.
They had noted, that when discovering such topics using social annotation,
one finds few topics that are very popular and have many experts,
and large numbers of topics that are quite niche and have fewer of experts.
\citeN{bhattacharya-recsys14} ranked a user's topics of interest
by simply ranking the topics
by the number of experts
the user had followed on that topic.
However, the large variance
in the number of experts available for each topic to follow
leads to an inherent popularity bias in interest inference.
Thus, for accurately modeling a user's interest,
it is necessary to take into account the popularity factor.
\emph{Finally}, the sheer volume of tweets available for a user to see,
still makes it a computationally difficult task
to perform recommendations in realtime.
Thus, it is necessary to make careful trade-off between accuracy and speed
to build a system capable of utilizing the above opportunity.

The major contributions of this paper are five fold.
\begin{NewText}
\begin{itemize}
\item (i) Extending the approach of our prior work \cite{ghosh-sigir12},
we present in Section~\ref{ssec:expertise},
a methodology for inferring human interpretable topics
of expertise of Twitter users,
by using Twitter List based social annotations.
\item (ii) Section~\ref{ssec:interest} presents
the Topical Preferential Attachment model,
which we employ to infer human interpretable topics of interests
of Twitter users,
by modeling their following behavior with respect to topical experts.
\item (iii) Section~\ref{ssec:candidate-tweets} presents
the Topical Binary Independence Model,
which we utilize to compute relative relevance of candidate tweets
with respect to given topics,
by utilizing information about tweets of topical experts.
\item (iv) In Section~\ref{ssec:recommend},
we present a recommendation methodology
that utilizes the information about user's interests
and the knowledge of topics of tweets.
Further, in Section~\ref{sec:system},
we discuss the working
of a  web based recommendation system
that utilizes the above approach,
and can be used by any active Twitter user.
\item (v) We evaluate our system against matrix factorization
based collaborative filtering,
content based recommendation,
and social recommendation (Section~\ref{sec:eval}).
Our approach shows competitive performance
against these baselines
under user based evaluations with 55 users.
\end{itemize}
\end{NewText}

To quantify the efficacy of our interest inference
and recommendation methodology,
we performed a controlled user study.
Evaluators were recruited from the authors' university
via advertising in mailing lists.
In the end, we obtained judgments from 55 evaluators
on the quality of the interest inference
and the recommendation methodology.
The evaluators were asked to rate top 10 topics of
interest inferred using our methodology
and that of \cite{bhattacharya-recsys14}
on a five point Likert scale.
We found that, when comparing to the baseline,
our method performed better
with a mean average score of 3.627
compared to 3.567 for the baseline.
Further, we compared overall recommendation methodology
to that of traditional recommendation techniques
such as matrix factorization based collaborative filtering (FunkSVD),
social recommendation,
and content based recommendation.
We found that overall our topical recommendation methodology
outperformed the above three baselines
in terms of mean average score,
while slightly trailing behind social recommendation
in terms of mean nDCG.

\section{Related works}
\label{sec:related}


Growth of the internet
and proliferation of available information
has led to
an ever increasing information overload
and an ever increasing dilemma of choices.
To help users navigate through
this morass of choices,
developers of online systems
have traditionally used recommender systems.

\subsection{Recommender systems in general}

Of the different recommender system methodologies
the most popular have been
collaborative filtering~\cite{shi-csur14}
and content based systems~\cite{lops-book11}.
Large volumes of literature
have been published
on developing recommender systems
using the above two techniques,
and their variations.
In the current work, however,
we take a completely different approach;
first, using social annotations
to infer interests of users,
followed by finding content
that are on topics of their interest.
Thus, we do not discuss
collaborative filtering and content based approaches
in much detail here.
Readers interested in thorough surveys on these methods
are referred to works by \citeN{shi-csur14} and \citeN{lops-book11}
for further details.

\subsection{Social recommendation}

Online social networking sites have also been facing
the effects of information overload,
as they are becoming more and more popular.
Many studies have been published
trying to develop recommender systems
specialized for social networks
\cite{ma-cikm08,chen-chi10,yang-www11,ye-sigir12,forsati-tois14,jiang-tkde14,chaney-recsys15,jiang-tkde15}.
Most of the works trying to develop
social recommender systems,
try to incorporate user-user relationships,
such as those available for friendship or follower-followee relations,
into traditional collaborative filtering or content based models
\cite{chen-chi10,cui-sigir11,chen-sigir12,hong-sigir12,zhang-www13,forsati-tois14,jiang-tkde14}.
The general idea behind these systems
is to exploit homophily and trust relations between users,
and to use them to constrain the recommendation model.

\citeN{yang-www11} postulated the correlation between
the interests of users who were friends on Yahoo!~Pulse,
and presented a `friendship-interest propagation' model,
trying to utilize the friendship network for recommendation.
Similarly, \citeN{ye-sigir12} presented
the `social influenced selection' model
based on the idea that friends have similar interests,
and the intuition that if a person likes an item
the probability of their friends liking it
grows significantly higher.
\citeN{hong-sigir12} presented a hybrid recommendation model
by using collaborative filtering
along with explicit item features.
Using LinkedIn as their test social network,
they presented a learning algorithm that used
the learning to rank paradigm.
\citeN{forsati-tois14} presented a recommendation model
that in addition to traditional trust relationships
obtained from user relations,
incorporated distrust relations
obtained from web-of-trust and personal block-lists.
Through experiments on the Epinions dataset,
they showed that incorporating such information
significantly improved the results.
\citeN{jiang-tkde14} presented a method that
distinguished itself by incorporating social contextual information
in addition to the regular user trust relations.
The key idea behind their approach being
to incorporate influence between users
as separate from trust relationships,
to help the recommendation algorithm.
When compared to baseline
their method outperformed others
on Renren and Tencent Weibo data.
\citeN{chaney-recsys15} noted that user trust relations
may not always be good for recommendation,
and a trusted user may point to items
that are not liked by the user.
Thus, they factored the trust relation between users
to Poisson factorization (SPF),
``a probabilistic model that incorporates social network information
into a traditional factorization method''.
\citeN{jiang-tkde15} created a novel hybrid social network
by combining information from multiple social networks
covering different item domains.
By using a random walk based link prediction algorithm
they predicted user-item links (recommendations)
and showed that it outperformed existing methods.

The primary aspect
where our approach differs from the above works is that,
we distinguish between expertise and interests of users,
especially in directed social networks, such as Twitter.
Our intuition is that,
if a user $u$ follows another user $v$,
them $u$ is likely to be interested in consuming content
which is being produced by $v$,
presumably on topics in which $v$ has expertise.
This is fundamentally different from saying that,
$u$ follows $v$ indicates $u$ and $v$
are likely to have similar interests.
While the interests and expertise of $v$ are likely
to have overlaps,
$v$ is likely to have interests beyond her topics of expertise.
Further, as will be discussed in more detail later,
$v$ can have expertise on a number of topics,
only a subset of which is of interest to $u$.
Our approach takes this finer grained view
of expertise and interests of social network users
and utilizes it for recommendation.

\subsection{Recommendations for Twitter}

As Twitter has grown popular,
its users have also been facing
the problems associated with information overload.
Many works have tried to alleviate this problem
by trying to build better content recommendation systems for Twitter
\cite{chen-chi10,kim-icdm11,chen-sigir12,yan-acl12,pan-recsys13}.

\citeN{chen-chi10} in their influential work,
presented the idea of social recommendation on Twitter.
They presented a comparative study,
comparing a number of content based
tweet recommendation algorithms,
along with a social voting system.
They noted that out of the variations they developed,
the version that had maximum performance
matched the content of potential tweets
to the user's own posted tweets.
The authors also noted that social voting,
that is weighing potential candidates for recommendation
by the number of times
similar tweets have been posted,
in the user's neighborhood,
produced better results.
\citeN{chen-sigir12} presented a hybrid recommendation system,
that used binary ratings (whether a user retweeted a tweet or not),
content features, and author features
to build an extended collaborative filtering system.
Interestingly, the authors noted the data sparsity problem
and difficulty of using retweets as rating source,
as very few tweets are retweeted.
To work around this problem,
they considered individual terms in tweets as items,
rather than whole tweets,
for the recommendation process.
The final part of the recommendation process
involved computing most relevant terms
and recommending tweets containing those terms.
\citeN{yan-acl12} presented a heterogeneous graph based model,
for the recommendation process.
The authors created a graph consisting of three parts:
(i) the user-user follower-followee graph,
(ii) the tweet-tweet text similarity graph,
and (iii) the user-tweet authorship graph.
Further, they used Latent Dirichlet Allocation (LDA),
to model users interests as high level latent topics,
using the tweets posted by a single user as documents.
They proposed a `co-ranking framework'
which utilized node rankings in the three
subnetworks (obtained via PageRank)
and user's latent interests
for performing recommendations.
In a recent work,
\citeN{pan-recsys13} extend their earlier work~\cite{chen-sigir12}
and tried to combine
the modeling of information diffusion
and the process of generating recommendations.
The key intuition behind their method
is to incorporate the already known
diffusion characteristics of a tweet
(current cascade size, current cascade depth,
popularity of existing users who have retweeted, etc.)
in to the collaborative filtering model.

The current work also tries to build a recommendation system
for the Twitter microblogging network.
However, the major difference between this work
and all the other ones presented above,
is that we use social annotations as our primary information source,
as opposed tweet ratings (such as retweets and likes).
Social annotations are first used to infer expertise
of a large and diverse set of Twitter users.
Next the inferred expertise of the expert users
is used to infer interests of users who follow them,
as well as to infer likely topics of tweets.
The use of a completely different information source
allows us to bypass the problems
traditionally faced with rating dependent systems.
As noted by previous works
\cite{chen-sigir12,pan-recsys13}
most tweets are not retweeted.
This sparsity in retweet data imposes a fundamental limitation
on the accuracy of traditional rating based approaches,
which is avoided by our social annotation based approach.
Additionally, many of the works tried to model user interests
in terms of latent representations.
While, this work also infers user interests,
it infers explicit human interpretable topics of interests,
rather than non-interpretable latent interests like the above works.
This allows us not only to use them to do recommendations
but also to explain them.

\subsection{Inferring interests of Twitter users}

As mentioned earlier,
the recommendation process described in this work
tries to infer explicit human interpretable topics of interest
for its users,
as an intermediate step.
Here we discuss earlier works that
have also tried to infer
human interpretable high level user interests.

\citeN{michelson-and10} presented a method
to infer high level topics of interests
for Twitter users.
Their methodology involved
extracting entities from tweets
and matching them with Wikipedia articles.
This was followed by
a traversal of the Wikipedia category tree~\cite{ponzetto-aaai07}
to obtain high level Wikipedia categories
that best matched with the Wikipedia articles,
found in the previous step.
The authors noted the difficulty
in validating interests of Twitter users.
For their work, they chose to manually validate
the inferred topics of interests of only four users.
\citeN{kapanipathi-eswc14} presented a method based on ideas
similar to \citeN{michelson-and10}.
Using a third party service,
they mapped entities within tweets to Wikipedia articles.
Further, they computed a Wikipedia category tree
and tried to obtain subtrees that best captured
the Wikipedia articles obtained previously.
The subtree thus obtained is presented as the
`Hierarchical interest graph' of the user.

The primary difference
between the apporach taken by the current work,
to infer interpretable topics of interests,
and the works describe above is that,
this work uses social annotations to infer user interests
for a user's expert followings,
rather than from the tweets posted by the users.
According to a recent study,
44\% twitter users have never posted tweets~\cite{Never-tweeters}.
This fact severely limits the percentage of the Twitter population
for whom such interests can be inferred.
Further, even for the fraction of users who post tweets,
it is not necessary they post tweets on all or even most topics
that are of their interest.
Thus, the coverage of interests that can be obtained from
such methods is also limited.
In contrast, inferring the interests of users
by means of analyzing the topical experts they are following,
is more robust towards the sparsity of tweet data,
and allows for capture of a more diverse set of interests.

\subsection{Inferring topics of tweets}

As part of the methodology of recommending tweets of users
based on their interests,
this work presents a simple method for finding topics of tweets.
Here we discuss existing works which delve in similar tasks.

A number of works have focused on inferring topics of tweets
using topic modeling
\cite{ramage-icwsm10,mehrotra-sigir13}.
\citeN{ramage-icwsm10} noted that the small size of tweets,
provides some unique challenges for topic modeling.
To overcome the problems of data sparsity
when performing topic modeling with a single tweet as document,
they utilized the semi-supervised topic modeling method, Labeled-LDA.
The topics inferred by their model
had both labeled (supervised) and unlabeled (latent) topics.
For the labeled topics,
the authors chose topic properties
such as hashtags, presence of different type of emoticons,
and presence of social features such as user mentions, replies, etc.
The authors manually categorized the unlabeled learned latent topics,
into the five classes: substance, status, style, social, and others.
Similar to the previous work,
To address the problem of data sparsity
due to the small size of tweets,
\citeN{mehrotra-sigir13} compared several tweet pooling techniques
to create documents for topic modeling.
Using techniques for measuring goodness of unsupervised clustering,
they measured the resultant clusters
to comment on the goodness of the tweet pooling scheme.
They noted that compared to other techniques
(such as author based pooling, temporal pooling, etc.)
hashtag based pooling system
performed best.


Multiple works have also been done trying to match
tweets to topics on top of a well known topic ontology or taxonomy.
Although the primary purpose of the works done by
\citeN{michelson-and10} and \citeN{kapanipathi-eswc14}
was to find interests of Twitter users,
as a prerequisite to that,
they mapped entities of tweets on to the Wikipedia category graph.
\citeN{yang-kdd14} presented a detailed description
of a topic inference system developed by Twitter.
The authors present high level descriptions
of creating a topic taxonomy
(derived from the Open Directory Project and Wikipedia)
and a semi supervised system that
maps non-chatter tweets to the topic hierarchy.






In a preliminary earlier study,
we utilized social annotation data
to design a simple system for inferring interests of users
(Who-Likes-What~\cite{bhattacharya-recsys14}).
While the present work borrows intuitions from our previous work,
it is significantly different from the prior work in a number of aspects.
First, the Who-Likes-What system in~\cite{bhattacharya-recsys14} used
a very basic methodology to compute the relative
interests of a user on a given topic -- by
simply counting of the number of experts that the user was following on the given topic.
\begin{NewText}
In contrast, the methodology for inferring interests of users
proposed in the current work
first develops a model for explaining user behavior
with respect to following of experts.
Next, it utilizes an expectation maximization based
learning algorithm to infer the topics of interest of the user.
As will be discussed in further detail later,
this approach allows us to address the problem
of popularity bias in interests ---
whereby most users have very popular topics as their
top topics of interest.
It also allows for accounting for the fact that
when a user follows an expert with multiple topics of expertise,
she may actually be interested in only  a subset
of the expert's topic of expertise.
Additionally, while the Who-Likes-What system
utilized noisy tags obtained directly from Twitter List descriptions,
the present work takes a different and improved
approach for mining topics from social annotations --
by utilizing Wikipedia corpus for extracting meaningful topics
as opposed to noisy tags.
This is necessary in our case,
as the topics mined by our system are also utilized for
the process of explaining recommendations to the user.
\end{NewText}


\section{Methodology}
\label{sec:method}

\begin{figure}
\centering
\includegraphics[width=\linewidth]{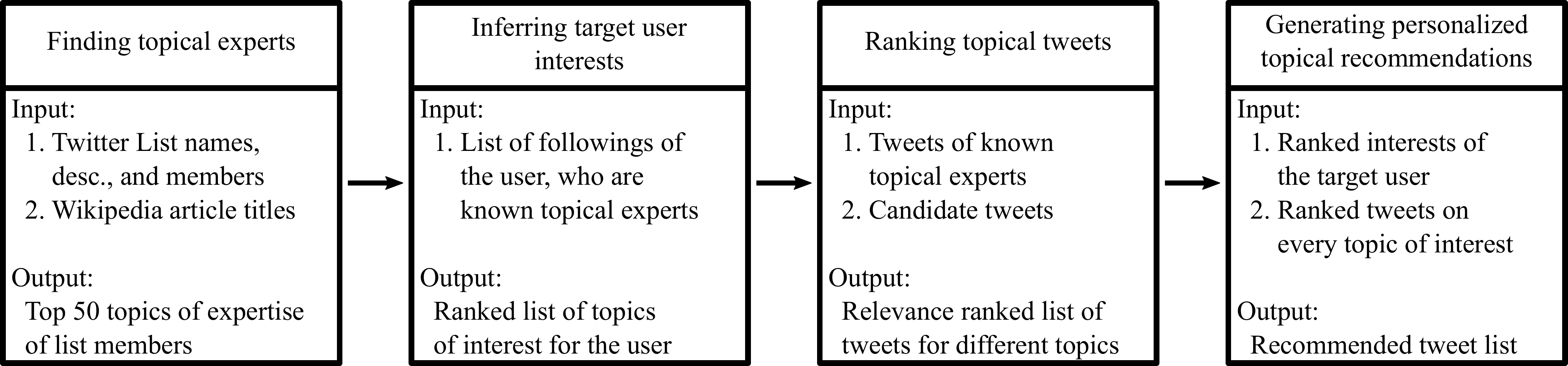}
\caption{An overview of the four overall stages
of the recommendation methodology proposed in the current work.}
\label{fig:methodology-block-diagram}
\end{figure}

\begin{NewText}
In this section, we describe our methodology
for generating personalized, topical tweet recommendations.
Figure~\ref{fig:methodology-block-diagram}
shows a high level overview of the four stage process
employed by our methodology.
\emph{First}, for a substantial and diverse set of topics,
we identify expert Twitter users,
who are well known for their expertise on these topics.
For this purpose, we utilize social annotations
mined from Twitter Lists.
\emph{Second}, for every target user,
for whom we generate recommendations,
we infer their relative interest in different topics.
This is accomplished by
modeling user behavior,
from her following of topical experts,
using a generative model,
that we call the Topical Preferential Attachment model.
We use an Expectation-Maximization (EM) algorithm~\cite{dempster-jrss77}
that uses the above generative model
to rank the target user's relative interest
in the various topics.
\emph{Third}, for a large corpus of tweets,
we compute relative relevance of every tweet
with respect to the different topics.
To efficiently compute relative relevance of
large volumes of tweets
with respect to many topics,
we propose a variant
of a probabilistic information retrieval model,
the Binary Independence Model (BIM)~\cite{lewis-ecml98},
that we call the Topical Binary Independence Model.
\emph{Finally}, having computed the relative interest
of a target user on different topics,
and having computed the relative relevance of
a large corpus of tweets with respect to
the various topics,
we employ a simple round robin algorithm
to generate a ranking of tweets,
with topical explanations,
for the target user.
The rest of this section details the steps presented above.
\end{NewText}

\subsection{Using social annotations to find topical experts}
\label{ssec:expertise}

\begin{table}
\begin{tabu} to \linewidth {lX[1l]X[1.8l]}
\toprule

Name
& Description
& Members\\

\midrule

neuroscience &
Neuroscientists and neuroscience related tweets &
@PsychToday,
@PsychCentral,
@OliverSacks (Neurologist),
@SamHarrisOrg (NeuroScientist) \\


recsys &
Recommender systems people &
@xamat (Xavier Amatriain),
@danielequercia,
@alansaid,
@wernergeyer \\

sociology &
Sociologists and sociological stuff &
@SocialPsych,
@bourdieu (French Sociologist),
@SociologyLens,
@markgammon (Sociologist) \\

pro-soccer &
Pro soccer players and clubs &
@FCBarcelona,
@Arsenal,
@10Ronaldinho (Ronaldinho Ga\'{u}cho),
@Cristiano (Cristiano Ronaldo) \\
\bottomrule
\end{tabu}
\caption{Example Twitter Lists: their names, descriptions, and sample members.%
\label{tab:list-examples}}
\end{table}

Our methodology, for generating topical tweet recommendations
relies extensively,
on having a large dataset of Twitter experts
with diverse topics of expertise.
For this purpose,
we extend the methodology
proposed in~\cite{sharma-wosn12,ghosh-sigir12},
that demonstrated the utility of social annotations,
derived from \emph{Twitter Lists},
in reliably inferring topics of expertise of popular Twitter users.

Lists are an organizational feature in Twitter
that allows users to create named groups of other Twitter users,
who are termed as members of the List.
For instance, a user can create a List named `Music and musicians' and 
add famous musicians like Eminem, Adele, and Lady Gaga to the List.
List creators can optionally add a more detailed description,
describing the members of the List.
Once a user has created a List,
she can then view the tweets
posted by the members of the List
separately from her regular tweet timeline.
Table~\ref{tab:list-examples} shows some example Lists,
created by different Twitter users,
along with some members of those Lists.

Users create Lists for their personal benefit,
that is to better organize the sources
from whom they want to receive tweets.
However, in doing so they give out strong signals
about the topical expertise of the List's members,
which can be mined from names and descriptions of the Lists.

\citeN{ghosh-sigir12}
built an expert search system
for finding experts on different topics,
based on tags mined from List names and description.
Although, the tags that were obtained were noisy
and in most cases did not represent meaningful topics,
they fulfilled the purpose of expert search,
given a query topic.
However, as the current work builds a recommendation system
that needs meaningful topics,
which are used for the purpose of explaining recommendations,
the above method is inadequate.

\begin{table}
\centering
\begin{tabu} to \linewidth {@{}ll@{}}
\toprule

Main Article
& Alternative Names \\

\midrule

Celebrity &
Celebs, Famous people, Media Personality \\

Television &
TV, Tely, TeeVee, T\'{e}l\'{e}vision \\

Sociology &
Social physics, Sociologist, Sosiology \\

United Kingdom &
UK, UKGBR, British State \\

Barack Obama &
Barack, Obama, Barry Obama, 0bama \\

\bottomrule
\end{tabu}
\caption{Examples of manually curated redirections in the English language Wikipedia.%
\label{tab:wiki-redirect}}
\end{table}

\textbf{Twitter List and Wikipedia dataset for finding experts:}
For the purpose of this work, we consider
\emph{article titles in the English language Wikipedia}
as the source of meaningful topics.
Specifically, for the results described in this work
we used the March 5, 2016, dump of the English Wikipedia corpus.
At a high level,
to find experts on a given topic $\Tpc$,
we first find all Twitter Lists in our database
that have the topic $\Tpc$ in their List name or description.
For this purpose, we obtained all Twitter Lists created
by the first 50 million Twitter users.
We were able to obtain 18.1 million distinct Lists
created by 7.4 million different users.

~\\
To create the corpus of topical experts
we use the following steps:

\noindent\textbf{Expertise inference step 1:
For every Twitter List find matching Wikipedia articles:}
To match Twitter Lists to Wikipedia articles,
we first convert the title of the articles into a query.
This is done by case folding the article's title
and removing any special characters.
Further, we ignore those queries that consist
entirely of known English, Spanish, and Portuguese stopwords,
and a custom set of Twitter specific stopwords,
such as Twitter, tweet, follow, etc.
Similarly, we normalize the List names and descriptions
by case folding the List names and descriptions
and removing any special characters.
Since List names can not exceed 25 characters,
multiple words in List names are often joined
using CamelCase.
Before the casefolding step,
we split List names that are in CamelCase
into individual words.
Finally, for every Wikipedia query
we search for the presence of the query
in the normalized List names and descriptions.

\noindent\textbf{Expertise inference step 2:
Merge articles titles that are substring of other titles:}
The query for the Wikipedia article `New York'
is a substring of the query for the Wikipedia article `New York Yankees'.
To ensure that most appropriate topics
are selected for a given List,
given such cases we only keep the longest matched title.
That is, if a List name or description matches two queries $q_1$ and $q_2$
and query $q_2$ is a substring of query $q_1$,
we discard shorter query $q_2$ for the given List.

\noindent\textbf{Expertise inference step 3:
Use Wikipedia redirects to normalize topic names:}
As List names and descriptions are freeform texts,
they contain a number of misspellings and abbreviations
of the actual topic.
To alleviate this problem,
we utilized Wikipedia redirects.
Wikipedia, contains a rich corpus of manually curated redirections,
that map commonly used abbreviations,
alternative wordings, and incorrect spellings
to the actual topic.
Table~\ref{tab:wiki-redirect} shows some sample redirections
present in the Wikipedia corpus.


\noindent\textbf{Expertise inference step 4: 
For all compute topical expertise:}
Finally, we compute the topical expertise of a user
by combining the user's List membership information,
the Twitter List to Wikipedia query mapping
(computed in steps 1 and 2),
and the Wikipedia query to Wikipedia article mapping
(computed in step 3).
As in \cite{ghosh-sigir12} we consider a user to be an expert on a given topic,
if he has been included in 10 Lists on the given topic.
Additionally, to limit the focus
on only the primary topics of expertise of a user,
we take only the top 50 topics of expertise for the given user.

\begin{table}
\begin{tabu} to \linewidth {lX[l]}
\toprule

Topic
& Experts\\

\midrule

Music &
@Eminem,
@Linkinpark,
@Beyonce,
@PinkFloyd,
@jtimberlake (Justin Timberlake),
@TaylorSwift13,
@LadyGaga,
@KatyPerry,
@KanyeWest,
@BrunoMars,
@Adele,
@springsteen (Bruce Springsteen) \\

Politics &
@BarackObama,
@HillaryClinton,
@realDonaldTrump,
@SenSanders (Bernie Sanders),
@JustinTrudeau,
@DWStweets (Debbie Wasserman Schultz),
@NarendraModi,
@David\_Cameron \\

Comedy &
@StephenAtHome (Stephen Colbert),
@ConanOBrien,
@TheEllenShow (Ellen DeGeneres),
@BillBurr,
@JimCarrey,
@JerrySeinfeld,
@KevinHart4real (Kevin Hart),
@iamjohnoliver (John Oliver) \\

Astronomy &
@MarsCuriosity (Curiosity Rover),
@TheRealBuzz (Buzz Aldrin),
@neiltyson (Neil deGrasse Tyson)
@NASAJPL,
@Cmdr\_Hadfield (Chris Hadfield),
@Space\_Station (Intl. Space Station) \\

\bottomrule
\end{tabu}
\caption{Example topical experts detected using the List based methodology.%
\label{tab:expert-examples}}
\end{table}

\begin{table}[t]
\centering
\begin{tabu} to \linewidth {X[l]X[2l]}
\toprule

List topics \cite{ghosh-sigir12} & Wikipedia filtered List topics \\

\midrule

fm, tv, celebs, bieber, new, noticias, bands, pr, os,
seo, sexy, melhores, jb, internet, foodies, bot, books,
medios 
&
Search engine optimization,
Real estate,
New York City,
Nonprofit organization,
National Football League,
Austin Texas,
Entrepreneurship,
Technology journalism,
The Walt Disney Company,
Veganism\\

\bottomrule
\end{tabu}
\caption{Sample of popular topics, generated using methodology
presented in \protect\cite{ghosh-sigir12}
and the Wikipedia based methodology.
Topics shown for each group are not present in the other group.%
\label{tab:topics-compare}}
\end{table}



Overall, using the above methodology,
we were able to find 741 thousand experts
on a diverse set of 11.6 thousand topics.
Table~\ref{tab:expert-examples} shows some sample topics
along with sample experts identified for those topics.

\textbf{Benefits of using Wikipedia for filtering List topics:}
To understand the difference between the topics identified
using our methodology and that of \cite{ghosh-sigir12},
we compare sample popular topics
identified using both, in Table~\ref{tab:topics-compare}.
A visible difference between our methodology
and that described by \cite{ghosh-sigir12} is that,
as our method does not have any limit on the number of terms
that can be part of the topics,
they are likely to be longer and more descriptive.
However, the main advantage of the Wikipedia based filtering methodology
is that topics identified using it
are more well formed and meaningful.

\subsection{Modeling expert following behavior for inferring interests of users}
\label{ssec:interest}

Traditionally collaborative filtering based recommender systems,
especially those based on matrix factorization,
have represented user interests using latent entities~\cite{shi-csur14}.
While such systems have been shown to produce
state-of-the-art results,
it can be quite difficult to interpret these latent interests,
and hence difficult to explain to users,
why they received a particular recommendation~\cite{shi-csur14}.
To address this issue, we propose a novel method for computing
\emph{human-interpretable} topics of interests of Twitter users.

\begin{NewText}
Our basic idea is to utilize
social signals that can be mined
from user behavior pertaining to following other expert users.
We utilize the basic intuition that
\emph{if a user follows a number of experts on a given topic,
the user is likely to be interested in that topic}.
However, to ascertain why a user is following another expert user,
from a single follow action alone is a difficult problem.
The expert user being followed
may have expertise on multiple topics,
making it difficult to determine
which of the multiple topics of expertise
a particular follower of her is interested in.
Further, many expert users are also popular celebrities,
and a follower of her may just be interested
about information related to her,
and not about any of her topics of expertise in particular.
Additionally, a user may also be a personal friend (in the offline world) of the expert user,
leading her to follow the expert,
without a general interest in the expert's topics of expertise.
Hence, to extract stable and meaningful information
about a user's interest from her follow actions,
we pose the problem of inferring interests of users,
from their expert followings,
as a problem of jointly modeling all her expert followings.
\end{NewText}

\noindent \textbf{The Topical Preferential Attachment model:}
At its heart, our method for modeling user follow behavior
is a generalization of the Preferential Attachment model~\cite{barabasi-science99},
that we call the Topical Preferential Attachment model.
We assume that for a user
the process of following an expert
is a two step process.
In the first step,
the user selects one of her topics of interest.
In the second step,
she preferentially follows an expert on that topic
based on the expert's popularity in that particular topical community.
That is, the more popular an expert is in her topical community,
the more likely that a user,
interested in the topic,
will follow the expert.
We assume that \emph{the number of times an expert has been
included in Lists related to that particular topic,
is indicative of her popularity in that topical community.}
However, there may be cases
where a user just wants to follow popular experts,
who are also well known celebrities,
without having any particular interest in their expertise.
To account for such cases,
we assume that an unnamed global topic exists,
in which all experts have expertise.
Further, similar to the Preferential Attachment model,
we assume that \emph{the follower count of the experts,
is indicative of how popular they are,
with respect to this global topic}.

Here we describe the generative model
that we utilize for jointly modeling a user's expert followings,
and present an Expectation-Maximization (EM) algorithm
to compute interests in realtime.

Let $\Tpcs' = \Tpcs \cup \{\Tpc_g\}$ be the set of all topics.
Here, $\Tpcs$ is the set of all regular topics
such as news, politics, music, etc.
Additionally, we define the global topic $\Tpc_g$,
different from regular topics,
as a special topic in which all experts have expertise.
Let the probability that a user is interested
in a certain topic $\Tpc$ be $P(\Tpc; \Int) = \IntTpc$,
where $\Int$ the corresponding stacked vector
of the probabilities $\IntTpc$ over all the topics.
Then,
\[
\forall_{\Tpc \in \Tpcs'} ~ \IntTpc \geq 0 %
~ \text{and} %
~ \sum_{\Tpc \in \Tpcs'} \IntTpc = 1
\]

Let $\Exps$ be the set of all experts.
Also, let the probability that an expert $\Exp \in \Exps$
is selected be $P(\Exp|\Tpc) = \TpcExp_\Exp$,
when a user is selecting an expert on topic $\Tpc$ to follow.
Additionally, $\TpcExp$ is defined as the corresponding stacked vector
of the probabilities $\TpcExp_\Exp$ over all the experts,
for the given topic $\Tpc$. Then, given a particular topic $\Tpc$,
\[
\forall_{\Exp \in \Exps} ~ \TpcExp_\Exp \geq 0 %
~ \text{and} %
~ \sum_{\Exp \in \Exps} \TpcExp_\Exp = 1
\]

The generative process describing
how a user selects her expert followings is as follows:
\begin{enumerate}[nosep]
\item First, the user selects a random topic $\Tpc$
using her own interest vector $\Int$~
following: $\Tpc \sim \Multinomial(\Int)$
\item Next, the user selects a random expert $\Exp$, on the topic $\Tpc$,
using the topic's popularity vector $\TpcExp$\\
following: $\Exp \sim \Multinomial(\TpcExp)$
\end{enumerate}

For regular topics $\Tpc \in \Tpcs$,
we define $\TpcExp_\Exp \propto \Listed_{\Exp,\Tpc}$,
where $\Listed_{\Exp,\Tpc}$ is the number of times
the expert $\Exp$ has been included
in a List on topic $\Tpc$.
For the global topic $\Tpc = \Tpc_g$,
we define $\theta^{\Tpc_g}_\Exp \propto \Followers_\Exp$,
where $\Followers_\Exp$ is the number of followers the expert has.

\begin{NewText}
\noindent \textbf{Using Expectation-Maximization for parameter inference:}
Next, we present an Expectation-Maximization algorithm
to infer the interest vector of a user $\Int$,
from her expert followings.
Expectation-Maximization is a class of iterative algorithms
for inferring model parameters
that uses the maximum likelihood method~\cite{dempster-jrss77}.
It is particularly useful for inferring model parameters
in cases involving unobserved latent variables.
In our model, the topics selected by the user,
from her interest vector,
in the first step of the generative model
are latent variables.
This makes the Expectation-Maximization approach
particularly suitable for our case.

Individual iteration steps
of Expectation-Maximization algorithms
are divided into two parts.
In the first part, known as expectation step or E-step,
the probability distribution of the latent variables are computed,
given the current estimate of model parameters.
In the second part, also known as maximization step or M-step,
maximum likelihood estimates of model parameters are computed,
given the observed data
and the probability distribution over the latent variables
computed in the previous step.
The iterations are continued until convergence of model parameters.
\end{NewText}

We now describe the Expectation-Maximization algorithm in detail.
Let the $\ExpFols \subseteq \Exps$ be the set of experts
that a user is following.
Further, let $\Tpcs^\Exp \subseteq \Tpcs'$ be the topics of
expertise of a given expert $\Exp$.
Given information about the set of experts $\ExpFols$
a user is following
we compute the maximum likelihood estimate of the user's interest, $\Int$.

The joint log likelihood function can be written as:
\begin{align*}
\ell(\Int)
  &= \sum_{\Exp \in \ExpFols}
     \log P(\Exp ; \Int)
   = \sum_{\Exp \in \ExpFols}
     \log
     \sum_{\Tpc \in \Tpcs'}
     P(\Exp | \Tpc) P(\Tpc ; \Int) \\
  &= \sum_{\Exp \in \ExpFols}
     \log
     \sum_{\Tpc \in \Tpcs^\Exp}
     P(\Exp | \Tpc) P(\Tpc ; \Int)
     \quad
     \because P(\Exp | \Tpc) = 0~\text{if}~\Tpc \not\in \Tpcs^\Exp
\end{align*}

Using the above log likelihood function
we can derive the E-step weights as,\\
\textbf{E-step:}
\begin{equation}
\wExpTpc
  = P(\Tpc|\Exp;\Int)
  = \frac{P(\Exp,\Tpc; \Int)}{P(\Exp; \Int)}
  = \frac{\TpcExp_\Exp \cdot \IntTpc}
         {\sum_{\TpcInTpcsExp} \TpcExp_\Exp \cdot \IntTpc}
\label{eq:estep}
\end{equation}

To obtain the M-step updates, we maximize:
\begin{align*}
\Int &= \argmax_{\Int}
        \sum_{\ExpInExpFols}
        \sum_{\TpcInTpcsExp}
        \wExpTpc
        \log
          \frac{P(\Exp,\Tpc;\Int)}{\wExpTpc} \\
     &= \argmax_{\Int}
        \sum_{\ExpInExpFols}
        \sum_{\TpcInTpcsExp}
        \wExpTpc
        \log
          \frac{\TpcExp_\Exp \cdot \IntTpc}{\wExpTpc}
\end{align*}
However, as $\IntTpc$ are probabilities,
the above maximization
is subject to the constraint $\sum_\Tpc \Int_\Tpc = 1$.
Thus, we construct and maximize the Lagrangian:
\begin{align*}
\mathcal{L}(\Int) =
  \sum_{\ExpInExpFols}
  \sum_{\TpcInTpcsExp}
  \wExpTpc
  \log
    \frac{\TpcExp_\Exp \cdot \IntTpc}{\wExpTpc}
  + \beta \left( \sum_{\Tpc \in \Tpcs'} \IntTpc - 1 \right)
\end{align*}
where $\beta$ is the Lagrangian multiplier.
Taking derivative and equating to zero we find:
\begin{align*}
\frac{\partial}{\partial \IntTpc} \mathcal{L}(\Int)
  = \sum_{\ExpInExpFols}
    \frac{\wExpTpc}{\IntTpc}
    + \beta = 0
\end{align*}
\[\text{or}\quad \IntTpc = - \frac{1}{\beta} \sum_{\ExpInExpFols} \wExpTpc \]
Since, $\sum_\Tpc \IntTpc = 1$ we have:
\begin{align*}
  \sum_{\Tpc \in \Tpcs'} \IntTpc
  = - \frac{1}{\beta}
    \sum_{\Tpc \in \Tpcs'}
    \sum_{\ExpInExpFols} \wExpTpc
  = - \frac{1}{\beta}
    \sum_{\ExpInExpFols} 1
  = - \frac{\left| \ExpFols \right|}{\beta} = 1
\end{align*}
\[\text{or}\quad \beta = - \left| \ExpFols \right| \]
Thus we have the M-step update rule,\\
\textbf{M-step:}
\begin{equation}
\IntTpc = \frac{1}{\left| \ExpFols \right|}
   \sum_{\ExpInExpFols} \wExpTpc
\label{eq:mstep}
\end{equation}

For our algorithm,
we initialize the vector of user interests $\Int$
in a data driven fashion,
instead of using a random initial value.
We choose the initial value of the
interest vector of a user $\IntTpc$,
to be proportional to
the number of experts on the topic
that the user is following.
For the global topic
we choose initial value of $i^u_{\Tpc_g}$,
to be proportional to the total number of experts
that the user is following.
Similar to \cite{carson-tpami02},
we find that initializing the system
in deterministic data driven manner
leads to the algorithm converging quicker
to more meaningful results.

Similar to \cite{carson-tpami02},
we run the update steps of the expectation maximization algorithm
is run till the relative improvement in subsequent
iterations of the algorithm falls below 1\%.


\begin{NewText}
\noindent \textbf{Benefits of inference based on followership modeling 
over expert counting:} The methodology for inferring topics
of interest presented above has a number of benefits
compared to the simple expert-counting methodology used
by the Who-Likes-What system in our prior work~\cite{bhattacharya-recsys14}.
\emph{First}, the Who-Likes-What system computed the relative
interests of a user on a given topic by
simply counting the number of experts that the user was following
on the given topic.
The major problem with this approach is that,
it fails to account for the popularity bias of topics.
A popular topic such as `Politics'
by virtue of being popular has many more experts on the topic,
than a niche topic such as `Neurology'.
Thus, it is intuitive that a user following equal number of experts
on both topics is likely to be more interested in the more niche topic.
But by simply counting the number of experts,
Who-Likes-What fails to account for the above phenomenon.
By using the modeling based approach presented above,
our method is accurately able to tackle this issue.
\emph{Second},
if a user follows an expert with multiple topics of expertise,
it is not possible for a simple expert counting algorithm
to infer which of the expert's topics
made the user follow the expert.
By taking into account the expert's relative popularity
in the different topical communities,
our modeling based approach is able to weigh the probabilities
of topical interest for the user.
\emph{Third}, an expert counting algorithm in general ignores
the relationship between topical tags.
For instance, a user who is an expert on `Libertarian Politics' in the USA
is likely to be also an expert on the general topic of `politics'.
Our inference algorithm is able to account for these topical interactions,
by utilizing a generative model of
following behavior.

To demonstrate the improvement in quality of interest inference
from our prior work~\cite{bhattacharya-recsys14},
we show in Table~\ref{tab:interest-examples},
the inferred interests of some sample users
using both the expert counting methodology in~\cite{bhattacharya-recsys14},
and the inference based method proposed in the current work.
The sample users shown in Table~\ref{tab:interest-examples}
are evaluators who participated in our human judgment
based evaluation process for judging the quality
of interest inference and recommendation presented
in the current work.
Detailed demographic information about the evaluators
who participated in our study is given later
in Section~\ref{ssec:eval-demographics}.

In can be clearly seen, that whereas the expert counting
based method is affected by popularity bias --
inferring highly popular topics such as, News, Tech, and Media 
as interests of most users --
the methodology proposed in this work does not suffer from such issues.
Additionally, our method is able to
extract meaningful niche topics --
such as, Computer science, Journalism,
Machine Learning, Public relations, and so on --
as relevant topics of interests for the given users.

\begin{table}
\begin{NewText}
\begin{tabu} to \linewidth {lX[l]X[1.2l]}
\toprule

Evaluator &
Interests inferred using expert counting
method of~\cite{bhattacharya-recsys14}
& Interests inferred using EM based inference
alogorithm proposed in current work \\

\midrule

P1 &
News, Social media, Media, Tech, Organization, Blog, Business,
Education, Gadget, Geek &
Web search engine, Ted, Journalism school, Mass media, Labs, Computer science,
Google News, Google products, Search engine marketing, Journalism \\

P12 &
News, Tech, Technology, Media, Science, Business, Blog, Information,
Organization, Technology journalism &
Google, India, University, Computer science, Machine learning, Nvidia,
Technical, Coding, Google Shopping, Mathematics \\

P17 &
News, Humour, Celebrity, Entertainment, Famous, Actor, Media, Hollywood,
Artist, Music &
Humour, Desi, Ubuntu, Director, Influence, Java, Google, Metallica,
Eminem, Breaking Bad \\


P33 &
Science, News, Research, Tech, Media, Blog, Business, Education,
Social media, Geek &
Information technology, Journal, Computer science, Machine learning,
Influence, Doctor of Philosophy, Research, Labs, NLP, CNET \\

P36 &
News, Media, India, Celebrity, Business, Journalist, Blog, Tech,
Entertainment, Politics &
India, Public relations, Bollywood, India News, Tech, Business journalism,
Global News, Corporate finance, Humour, News \\

\bottomrule
\end{tabu}
\end{NewText}
\caption{Top ten topics of interests,
of sample evaluators, 
identified using the simple expert counting based methodology
presented in~\protect\cite{bhattacharya-recsys14},
and the Expectation-Maximization based
inference algorithm presented in this paper.
Details about the evaluators,
for whom interests are shown,
is presented later in Section~\protect\ref{ssec:eval-demographics}.
\label{tab:interest-examples}}
\end{table}

\end{NewText}

\subsection{Selecting candidate tweets for recommendation process}
\label{ssec:candidate-tweets}

Having inferred the relative interests of users
in different topics,
the next of our recommendation process
involves finding important and interesting tweets
to recommend to users
related to their topics of interest.

While it would certainly be desirable
to compute relevance of all tweets posted in Twitter
to different topics,
with hundreds of millions of tweets posted every day,
it is a practical challenge to process all tweets in realtime.
Thus, many analytic companies and researchers are increasingly
resorting to some form of sampling.
In our prior work~\cite{zafar-tweb15},
we compared two orthogonal sampling strategies for sampling tweets:
random sampling and expert sampling.
The first method involved using the Twitter provided 1\% random sample,
while the second consisted all tweets
posted by over 500 thousand experts on a diverse set of topics.

It was noted that a number of properties of the expert sample
make it more desirable for information mining.
We had observed that tweets contained in the expert sample
contained much less spam, less chatter,
and had a much larger fraction of informative content.
Interestingly, we also found that,
although the tweets in the expert sample were obtained
from a very small fraction of the Twitter population,
they were a good representation of the overall crowd sentiment
on different events and entities.
Further, for a number of major news events,
the information about the event appeared in the expert sample
before it appeared in the Twitter streaming API.
The above features make expert sampling
a very attractive choice for the purpose of
tweet recommendation.

For creating the expert sample for the current work,
we regularly crawl new tweets,
posted by the 741 thousand experts in our dataset,
every 15 minutes.
During the month of April 2016
our expert sample
contained an average of 2.5 million tweets per day.

\subsection{Computing relevance of tweets with respect to different topics}
\label{ssec:twttopic}

Although our recommendation process
considers candidate tweets that are tweeted
by experts with known topics of expertise,
it is still necessary to ascertain
the relative relevance of the chosen tweets
to the given topics, for a number of reasons.
First, it is not necessary for users who are experts on specific topics
to limit their tweets within the boundaries of their topical expertise.
In our previous work~\cite{zafar-tweb15},
we actually observed that, apart from posting tweets relevant
to their expertise, the topical experts often participate in day-to-day conversations,
and tweet about topics beyond their topics of expertise.
Second, many topics with broad public appeal,
such as sports, entertainment and politics, are discussed by everyone,
including users who are not experts on those subjects.
Further, within our dataset,
a user may have multiple topics of expertise.
Thus, it is important to ascertain
which topic a tweet is about,
when an expert with a multitude of expertise
posts a tweet.
Finally, even if the topic of a tweet is explicitly known,
it is important for the purposes of recommendation
to obtain a relative ranking for the tweets on a given topic.

For this work, we take a probabilistic approach
in trying to ascertain the relative relevance of a tweet
with respect to a given topic.
Ideally, we want to answer the following question,
for every tweet-topic pair:
what is the probability that the given tweet
is on the given topic?
We try to approximate the answer to the above question
with the answer of the following one:
given information only about the content of the tweet,
what can we tell about the probability,
that the given tweet was posted by an expert on the given topic?
The intuition behind the above approach is that,
if the contents of a tweet make it seem,
that an expert on the given topic had tweeted it,
it is likely that the tweet itself is relevant to the topic.

The problem of computing
the likelihood of a tweet being posted by an expert on a given topic,
from the content of the tweet,
implores the use of a classification framework.
Further, given the dataset of topical experts and their tweets,
it is quite straight forward to set up
one of the common state-of-the-art
text based classification frameworks,
such as Linear SVM or Naive Bayes Classifier~\cite{schulz-emnlp15}.
However, given that fact that our dataset consists of
over 11.6 thousand different topics,
one would have to train at-least 11.6 thousand
different classifiers in one-vs-all mode.
To avoid the expensive operation of training
such a large number classifiers,
we take an alternate approach
based on the Binary Independence Model (BIM)~\cite{lewis-ecml98}.

\noindent
\textbf{The Topical Binary Independence Model:}
The Binary Independence model or BIM
is a probabilistic document retrieval,
model based on the Naive Bayes Classifier.
The primary reason for using a BIM inspired model
rather than directly using Naive Bayes Classifier is that,
the vanilla BIM model makes few reasonable assumptions
over and above the `Naive Bayes' assumption
of the Naive Bayes Classifier,
that makes computation of the model parameters
much faster and storage efficient.
Additionally, the time complexity of computing the ranking score function,
of a tweet with respect to a topic,
is orders of magnitude faster,
when compared to the regular Naive Bayes Classifier.
However, this speedup in model training
and computing ranking score
comes at the cost of not being able
to compare the relevance of tweets across topics.
This is due to the fact that
the document rankings computed for different topics
using the BIM model,
are independent of each other.
However, we found this trade off acceptable for our purpose,
as having already computed the topics of interest of a given user,
our recommendation algorithm only needs
relative rankings of tweets with respect to given topics.

For computing tweet rankings with respect to different topics
using our Topical BIM model,
we represent a tweet as a bag of entities, $\Twt \subseteq \Ents$,
where an entity $\Ent \in \Ents$ is either present or absent in the tweet.
Here, $\Ents$ is the set of all entities in our corpus.
To extract meaningful entities from tweets we used the following steps.
First, we filtered out non-English tweets from our dataset.
For this purpose we used the language field provided by Twitter.
Any tweet that did not have its language field set as `en', were removed.
Next, we processed the tweets using
the CMU ARK part-of-speech Tagger~\cite{owoputi-naacl13}.
We selected as entities,
terms marked with the following POS tags:
proper nouns, common nouns, and \#hashtags.

We denote, relevance of a tweet $\Twt$ to the topic $\Tpc$
using $P(\Tpc | \Twt)$,
which represents the probability
the tweet's author is an expert on topic $\Tpc$
given the content of the tweet.
Ranking tweets using the above probability
is equivalent to
ranking them using the following ranking function:
\begin{align}
\fRank(\Twt, \Tpc) %
  &= \OddsRatio P(\Tpc | \Twt) %
   = \frac{P(\Tpc | \Twt)} %
         {P(\bar{\Tpc} | \Twt)} \\
  &= \frac{P(\Twt | \Tpc) \cdot P(\Tpc) / P(\Twt)} %
         {P(\Twt | \bar{\Tpc}) \cdot P(\bar{\Tpc}) / P(\Twt)} \\
  &= \frac{P(\Twt | \Tpc)}{P(\Twt | \bar{\Tpc})}
     \cdot
     \frac{P(\Tpc)}{P(\bar{\Tpc})}
\end{align}
Here, $P(\bar{\Tpc} | \Twt)$ represents the probability
that the tweet's author is not an expert on topic $\Tpc$.
As the term $P(\Tpc) / P(\bar{\Tpc})$ is a constant
when ranking relevant tweets for a given topic,
the ranking function can be further simplified to:
\begin{align}
  \fRank(\Twt, \Tpc) = \frac{P(\Twt | \Tpc)}{P(\Twt | \bar{\Tpc})}
\end{align}
Next, we make the `Naive Bayes' conditional independence assumption:
the probability of presence of different entities in a tweet
are independent of each other,
given the topic of expertise of the tweet's author.
Then the above ranking function can be written as,
\begin{align}
\fRank(\Twt, \Tpc) %
  = \prod_{\Ent \in \Ents} \frac{P(\Ent | \Tpc)}{P(\Ent | \bar{\Tpc})} %
  = \prod_{\Ent \in \Twt} \frac{\pEntTpc}{\qEntTpc}
          \cdot %
    \prod_{\Ent \not\in \Twt} \frac{(1 - \pEntTpc)}{(1 - \qEntTpc)}
  \label{eq:bim-score-1}
\end{align}
Where $\pEntTpc$ is the probability
that $\Ent$ occurs in a tweet
posted by an expert on topic $\Tpc$,
and $\qEntTpc$ is the probability
that $\Ent$ occurs in a tweet
not posted by an expert on topic $\Tpc$.
Next, we make the first BIM model assumption,
also known as the \emph{relevance of empty document assumption}:
that is a tweet with no entities present in it
is equally probable to be tweeted by expert on topic $\Tpc$
and a non-expert on topic $\Tpc$~\cite{lewis-ecml98}.
This allows us to write the ranking function as:
\begin{align}
\fRank(\Twt, \Tpc) %
  &= \prod_{\Ent \in \Twt} \frac{\pEntTpc}{\qEntTpc} %
     \cdot %
     \prod_{\Ent \not\in \Twt} \frac{(1 - \pEntTpc)}{(1 - \qEntTpc)} %
     / %
     \prod_{\Ent \in \Ents} \frac{(1 - \pEntTpc)}{(1 - \qEntTpc)} \\
  &= \prod_{\Ent \in \Twt} %
     \frac{\pEntTpc (1 - \qEntTpc)}{\qEntTpc (1 - \pEntTpc)}
  \label{eq:bim-score-2}
\end{align}
Here,
$\prod_{\Ent \in \Ents} (1 - \pEntTpc)/(1 - \qEntTpc) %
= P(\emptyset | \Tpc) / P(\emptyset | \bar{\Tpc}) = 1$
is the ratio of two probabilities:
(i) the probability that the empty tweet
was posted by an expert on topic $\Tpc$, $P(\emptyset | \Tpc)$,
and (ii) the probability that the empty tweet
was not posted by an expert,
whose expertise did not contain topic $\Tpc$, $P(\emptyset | \bar{\Tpc})$.
The above ratio is equal to $1$ by the relevance of empty document assumption.

Having the expert sample, the probability $\pEntTpc$
can be estimated as $\nEntTpc/n_\Tpc$.
Here, $\nEntTpc$ is the number of tweets in the corpus
posted by experts on topic $\Tpc$,
that contained the entity $\Ent$,
and $n_\Tpc$ is total number of tweets posted by experts on topic $\Tpc$.
As $\qEntTpc$ represents the number of tweets
not belonging to topic $\Tpc$
that contained entity $\Ent$,
the computation of $\qEntTpc$ results in a dense matrix,
which is difficult to store and operate on.
Hence, we make the second BIM model assumption:
for a given topic,
as most documents in the corpus have not been tweeted
by an expert on the topic,
the whole corpus of tweets is a good approximation of tweets
not tweeted by an expert on the specific topic.
Thus, we can approximate:
$\qEntTpc$ as $q_\Ent=n_\Ent/n$,
that is probability that the entity $\Ent$ occurs
in any given tweet in the corpus.
Here, $n_\Ent$ is the total number of tweets in the corpus
containing entity $\Ent$,
and $n$ is the total number of tweets in the corpus.
Thus, the ranking function can be further simplified to:
\begin{align}
\fRank(\Twt, \Tpc) %
  &= \prod_{\Ent \in \Twt} %
     \frac{\pEntTpc (1 - q_\Ent)}{q_\Ent (1 - \pEntTpc)} %
   = \prod_{\Ent \in \Twt} %
     \frac{\frac{\nEntTpc}{n_\Tpc} (1 - \frac{n_\Ent}{n})} %
         {\frac{n_\Ent}{n} (1 - \frac{\nEntTpc}{n_\Tpc})} \\
  & = \prod_{\Ent \in \Twt} %
     \frac{\nEntTpc (n - n_\Ent)}%
          {n_\Ent (n_\Tpc - \nEntTpc)}
\end{align}

To further speedup the computation of the score function at runtime,
and to guard against over fitting,
we rewrite the above equation as:
\begin{align}
\fRank(\Twt, \Tpc) = \prod_{\Ent \in \Twt} \sEntTpc
\end{align}
where
\[
  \sEntTpc =
  \begin{cases}
     \dfrac{\nEntTpc (n - n_\Ent + \delta)}{n_\Ent (n_\Tpc - \nEntTpc + \delta)}
     & \quad \text{if } \nEntTpc \geq k \\
     1
     & \quad \text{otherwise}  \\
  \end{cases}
\]
Here, $\delta$ is a positive smoothing factor
which also ensures that none of the terms
reduce to zero,
and $k$ is model parameter
that ensures that an entity occurs at-least $k$
times in tweets from experts on topic $\Tpc$
before it is used for rank calculation.
For the purposes of this work, we chose $\delta=1$ and $k=5$.
Finally, to guard against the potential bias of the ranking function
towards longer tweets with more entities,
we rewrite the final ranking function as:
\begin{align}
\fRank(\Twt, \Tpc) = \prod_{i \in \{1..\nu\}} s^{~i}_{\Ent,\Tpc}
\end{align}
where $s^{~i}_{\Ent,\Tpc}$ is the $i$th largest value
in the collection $\{\sEntTpc : \Ent \in \Twt\}$.
For this work, we experimented with a range of thresholds,
and found that value of $\nu=3$ works well in practice.

\begin{NewText}
\noindent
\textbf{Suitability of using BIM inspired model in current scenario:}
Although BIM has been an influential model,
it has been heavily criticized~\cite{lewis-ecml98} for some
assumptions it makes, especially for making strong assumptions about
probabilities $\pEntTpc$, ignoring term counts in documents,
the conditional Naive Bayes assumption, and ignoring document length.
Numerous works have been published since, trying to address the above limitations.
However, for our current purpose,
we find the BIM inspired model quite effective
as the controversial assumptions do not
create significant issues
in the domain where we use it.
First, in the document retrieval setting,
where BIM model was originally used,
the model had to make assumptions about
the probabilities $\pEntTpc$,
as it was not possible to know a priori
which documents would be relevant
to an arbitrary query.
However, for our task,
since we already know which tweets
were made by which experts,
we can directly compute $\pEntTpc$.
Further, as tweets are small,
it is unlikely that relevant tweets
have much repetition of entities,
thus making the binary entity count assumption a reasonable one.
\citeN{cooper-tois95} noted that
the Naive Bayes assumption in the BIM model
(and hence our model)
can be replaced with a much weaker and more reasonable
``linked dependence assumption''
without changing the model computation.
This allows models using the Naive Bayes assumption,
such as our method, to perform much more competitively
than what would be expected
of models making such a drastic and simplistic assumption.
Additionally, the BIM inspired model used in the current work
does not ignore document lengths
and corrects for the problem
by limiting the effective number of entities
that are used in computing the final topical rank score of a tweet.

\noindent
\textbf{Source of speedup in Topical BIM:}
It straightforward to see
that training our Topical BIM model,
which needs only one pass over the data
to obtain frequency counts of the distinct entity-topic pairs,
is much faster than
training 11.6 thousand Linear SVM models
in one vs all mode,
each of which require an expensive
gradient descent based optimization schedule.
However, it is not so obvious to see the speedup
obtained from using Topical BIM
over a traditional Naive Bayes classifier,
which has similar training times.
The main speed advantage of Topical BIM
comes from utilizing
the `relevance of the empty document assumption',
which reduces the time complexity
of computing the relevance score of a tweet to a topic $\fRank(\Twt,\Tpc)$,
from the order of total number of entities in {\it the corpus},
to the order of the number of entities in {\it the tweet}.
This change can be seen,
by observing how the computation
of the  scoring function $\fRank(\Twt,\Tpc)$,
changes from Eq.~\ref{eq:bim-score-1} to Eq.~\ref{eq:bim-score-2}.
To compute the score function in Eq.~\ref{eq:bim-score-1}
one would have to compute the product of $|\Ents|$
separate terms.
While after using the `relevance of the empty document assumption'
the number of terms required to compute the score function
in Eq.~\ref{eq:bim-score-2}
reduces to $|\Twt|$.
The median number of entities present in the daily tweet corpus,
posted by 741 thousand experts in our dataset,
during the month of April 2016,
came to be 363,553.
While the median number of entities present in individual tweets,
for the same dataset,
came to be 5.
This represents a 72.7K (thousand) times
increase in tweet-topic scoring speed
that is obtained from using the Topical BIM model,
compared to a vanilla Naive Bayes classifier!

The rate of decrease in the cost of computation power
due to advancement in computational hardware
is clearly lower
than the ever increasing pace at which
content is being produced in online social media today.
Thus, it is imperative to be able to judiciously
trade off computation speed and computation accuracy,
to be able to process millions of pieces of content in real-time.
The efficiency of the Topical BIM model presented above,
significantly contributed in enabling us to build a practical system,
that we describe in Section~\ref{sec:system},
for demonstrating the efficacy of our approach.
\end{NewText}

\subsection{Recommending tweets to users}
\label{ssec:recommend}

In Section~\ref{ssec:interest}
we described the Topical Preferential Attachment model
for inferring relative interest of Twitter users in different topics.
This was followed by Section~\ref{ssec:twttopic}
which described the scalable Topical Binary Independence model
for scoring the relative relevance of a tweet
with respect to given topics.
In this section, we present the overall recommendation algorithm
that utilizes the above methods
to create personalized tweet ranking
to be presented to the user.

The steps of the overall recommendation algorithm
are as follows:
\begin{itemize}
\item \textbf{Step 1:} For the target user $u$ for whom the recommendations
  are to be generated,
  compute the top $M$ interests of the user
  using the Expectation-Maximization algorithm
  described in Section~\ref{ssec:interest}.
\item \textbf{Step 2:} For each of the $M$ topics of interest
  obtained in previous step,
  compute the top $N$ most relevant tweets
  using the Topical BIM model
  described in Section~\ref{ssec:twttopic}.
\item \textbf{Step 3:}
  For each of the $M$ ranked lists of tweets,
  corresponding to the $M$ topics of interest,
  remove tweets that are similar to other tweets in the list.
  If two similar tweets
  with rank $r_1$ and $r_2$ are present in a single list,
  and $r_1 < r_2$ then remove the tweet with rank $r_2$ from the list.
\item \textbf{Step 4:} Fill up the recommendation list
  for the user $u$, by selecting tweets from the
  $M$ topical lists in a round robin fashion.
\item \textbf{Step 5:} From the final recommendation list
  created the in the last step,
  remove tweets that are similar to other tweets in the list.
  If two similar tweets
  with rank $r_1$ and $r_2$ are present in a single list,
  and $r_1 < r_2$ then remove the tweet with rank $r_2$ from the list.
\end{itemize}

For the above procedure
of generating the recommendation list of tweets,
we consider two tweets to be similar,
if the Jaccard similarity of their entity sets
is greater than or equal to 0.7.

For the evaluations described in the following section,
we choose the number of topics $M=50$ for every user,
and for every topic we compute the top $N=1000$ tweets.

\section{Evaluation}
\label{sec:eval}

In this section,
we try to quantify the efficacy
of our interest inference algorithm
and our tweet recommendation methodology.
Most recommendation studies
utilize an offline evaluation approach~\cite{chen-sigir12,pan-recsys13,jiang-tkde15}.
In these studies,
the recommendation algorithm produces
a ranking of recommended items,
for every given user.
The goodness of the recommendation algorithm is measured by
how high-ranked are the items,
\emph{that are already known to be liked by the user}.
However, as the tweets recommended by our system
are likely to come from sources,
that the user does not follow,
such as topical experts other than the ones
that the user is following,
it is unlikely for the users to have liked them,
making it difficult
to utilize the above evaluation approach.
Thus, to compare the recommendations generated
by our methodology
with existing works,
we take a user study based evaluation approach,
where human evaluators are asked
to rate the top recommendations
generated for them,
on a five point Likert scale.
This method of evaluation has an additional advantage for our case.
As an intermediate step,
our algorithm computes meaningful topics of interest,
for the given user.
Evaluating interest inference
is an inherently  difficult problem,
as only the users whose interests are inferred
can judge the quality of the results.
This makes a user-study based evaluation
particularly suitable for the current task.

\subsection{Demographics of evaluators}
\label{ssec:eval-demographics}

Active Twitter users
were recruited as evaluators,
by advertising in the
mailing lists of the first author's institute.
The evaluators were given
a token compensation
for participating in the process.
As discussed earlier,
a large fraction of Twitter users
are passive consumers,
not tweeting, or liking many tweets.
This makes it difficult
to apply traditional rating based recommendation algorithms,
which use retweets and likes as ratings.
To ensure that our comparison with
the above baselines are fair,
we required our evaluators to have
followed at-least 50 users,
and have retweeted or liked at-least 10 tweets.
A total of 55 evaluators
participated in the study.
All evaluators were Indian nationals.
The age of the evaluators ranged between 19 and 28 years,
with the median being 23 years.
Six out of the 55 evaluators were female,
while the rest were male.
Of the 55 evaluators,
21 were graduate students,
16 were under-graduate students,
and 18 were employed in the industry.

\subsection{Evaluating interest inference}

Our interest inference algorithm,
infers high-level topics of interests for Twitter users.
This is in contrast with most earlier works
where inferred interests have one of the two forms:
(i) latent interests in an abstract domain~\cite{chen-sigir12,pan-recsys13}
or (ii) distributions of tags and terms computed from tweets,
received or posted by the Twitter users~\cite{chen-chi10}.
The interest inference apporach
that is most similar to the current work
is that of our earlier work \cite{bhattacharya-recsys14},
where we computed a user's relative interest in a topic
by simply counting the number of experts
that the user was following on that topic.

In this work, we compare the Expectation-Maximization based 
interest inference algorithm of our `What You Like' system (WYL-Infer),
with the expert counting based approach
of the `Who Likes What' system
presented in our prior work \cite{bhattacharya-recsys14} (WLW-Count)
because of the following reasons:
since, it is difficult for a human user
to understand the underlying meaning of abstract latent interests,
it is difficult for them to directly compare
those abstract latent interests,
with the meaningful topics of interests that our algorithm produces.
Additionally, we have already demonstrated
that even simple expert counting works better at extracting
meaningful interests of users
than tag distributions obtained
from using language modeling techniques
on posted or received tweets.

For our evaluation,
we used our dataset of user expertise,
to compute top 10 topics of interests
of every user who participated in the evaluation process,
using our EM based inference algorithm (WYL-Infer)
as well as expert counting method (WLW-Count).
For every topic,
the evaluators were asked to rate,
how much they were interested in that particular topic,
on five point Likert scale,
ranging from `Very interested' (score 5)
to `Not interested at all' (score 1).

To compare the results of the evaluation process
we measure overall relevance of the methodologies
using mean average score
(mean of the average scores computed per user)
and mean precision.
For computing precision,
which requires binary relevance scores,
we counted rating scores of 4 and 5 as relevant,
while scores between 1 and 3 were counted as irrelevant.
Additionally, we measured the goodness of the rankings
obtained from the different systems
using mean average precision (MAP) and
mean normalized discounted cumulative gain (nDCG).

\begin{table}
\begin{tabu} to \linewidth {@{}X[1.8l]X[r]X[r]X[r]X[r]@{}}

\toprule

\rowfont[c]{}

& Mean Average Score & Mean Precision & Mean Average Precision  & Mean nDCG \\

\midrule

WLW-Count   & 3.567          & 0.589          & 0.725          & 0.916          \\
WYL-Infer (proposed)   & \textbf{3.627} & \textbf{0.607} & \textbf{0.757} & \textbf{0.919} \\

\bottomrule
\end{tabu}
\caption{Performance comparison of
WLW-Count: the expert count based
interest inference strategy~\protect\cite{bhattacharya-recsys14},
and WYL-Infer: the proposed EM based interest inference algorithm.%
\label{tab:interest-eval}}
\end{table}



Table~\ref{tab:interest-eval}
shows the summary of the evaluation results.
We find that when compared to the expert counting based approach (WLW-Count)
the current EM based interest inference (WYL-Infer)
shows modest to marginal improvements across
all the chosen metrics.
The largest improvements are observed
for the binary precision based measures,
with 3.05\% improvement in terms of mean precision,
and 4.41\% improvement using mean average precision.
However, when comparing using absolute scores,
the improvements are smaller:
1.68\% improvement in terms of mean average score,
and 0.32\% improvement as per mean nDCG.

\begin{figure}
\centering
\subfloat[WYL-Infer]{\label{fig:interest_cloud_infer}%
    \includegraphics[width=0.49\textwidth]{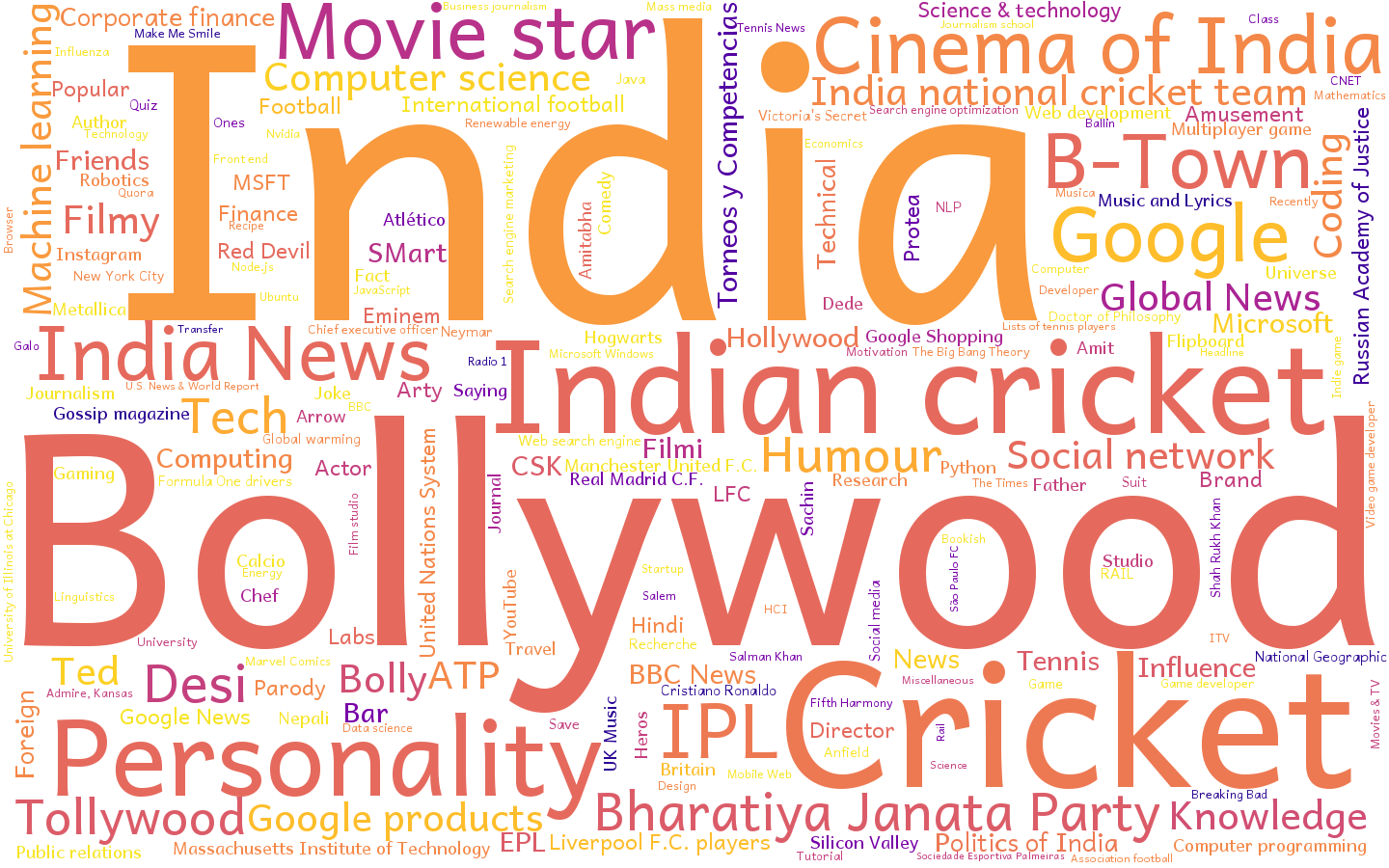}%
}%
\hfill
\subfloat[WLW-Count]{\label{fig:interest_cloud_raw}%
    \includegraphics[width=0.49\textwidth]{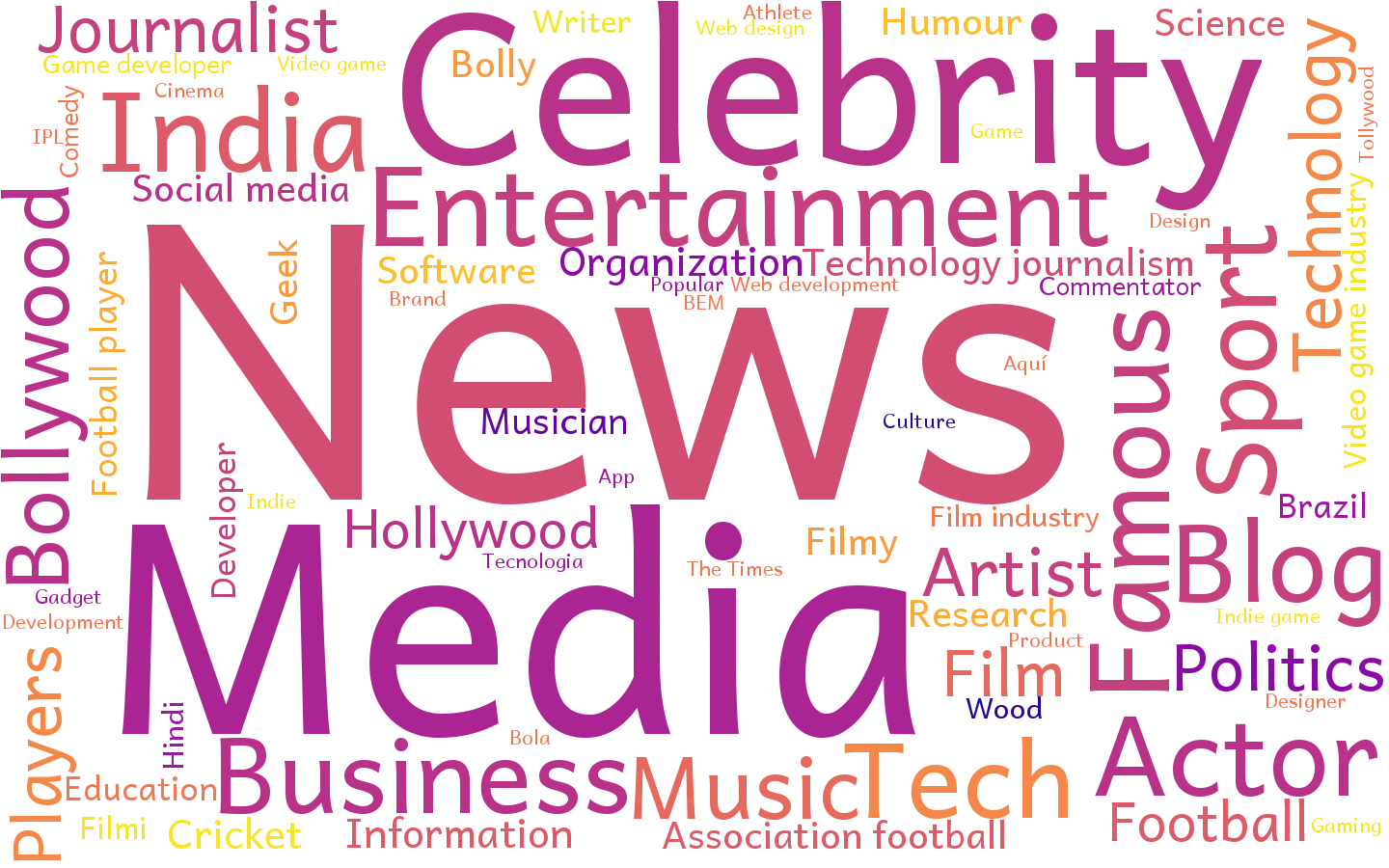}%
}%
\caption{Comparing the topic clouds
created from the top 10 interests of the 55 evaluators,
computed using the interest inference method (WYL-Infer)
and expert counting method (WLW-Count).
Font size of the topic in the wordclouds
is proportional to the number of evaluators
for whom the topic was inferred as an interest.
The brightness of the font's color
is proportional to the average interest score
received by the topic (brighter is better).}
\label{fig:interest_cloud}
\end{figure}

Next, to understand the qualitative difference
in the results obtained from the two methods,
we compare the topic clouds
of the 10 topics of interests,
generated for the 55 evaluators
using the two methods.
Figure~\ref{fig:interest_cloud} shows
generated topic clouds.
The font size used to depict a topic
is proportional to the number of evaluators
for whom the topic was inferred as an interest
while the brightness of the topic
is proportional to the mean score
received by that topic.

It is easy to observe,
that the diversity and granularity
of topics inferred using the proposed
method is much higher.
For the 55 evaluators,
the top 10 topics of interests
inferred by the inference method (WYL-Infer)
yielded  237 distinct topics.
While the same for the expert counting method (WLW-Count)
was only 73.
Additionally, the topics inferred by the
inference method better captures the
local interests of the Indian evaluators:
India News, Indian Cricket,
Indian Cinema (Bollywood, Tollywood, Cinema of India),
Indian political parties (Bhartiaya Janta Party).
While those inferred by the expert counting method (WLW-Count)
are much more generic in nature,
like music, movies, politics, etc.


Interestingly, we find that topics that are inferred as interests for many evaluators,
by the expert counting method (WLW-Count),
are likely to be more disliked by the evaluators.
The Spearman rank correlation between the
mean score obtained by a topic of interest
and the number of evaluators for whom the topic was inferred as interest
is $\rho=-0.290$ ($p=0.013$).
However, similar correlation does not exist for the
topics inferred by our current inference based method (WYL-Infer):
$\rho=-0.064$ ($p=0.321$).
This can also be seen in Figure~\ref{fig:interest_cloud}
where the larger topics in Figure~\ref{fig:interest_cloud_raw} are darker,
than the larger topics in Figure~\ref{fig:interest_cloud_infer}.

To better explain the cases,
where our inference based method (WYL-Infer) was performing poorly,
we followed up with the evaluators to understand
why they found particular inferred topics of interests uninteresting.
We found in general the cases followed the following themes.

\emph{Foreign topic descriptions were more disliked}:
Even though the topics inferred by our methodology
are limited to topics present in English language Wikipedia,
our topical dataset contains a number of non-English topics.
This is because the English language Wikipedia contains
a number of topics with non-English names.
In particular, we found that one of the evaluators
was interested in a number of Argentine Footballers,
because of which our algorithm found him to
be interested in `Torneos y Competencias'
an Argentine sports firm.
However, the evaluator was not acquainted in this particular institution
and found it not interesting.
Similarly, we found another evaluator who is interested
in the South African Cricket Team,
who are nicknamed `the Proteas'.
However, the evaluator was not familiar with this term
and marked the topic uninteresting.
Additionally, we found evaluators who were following
a number of popular performers from United Kingdom,
but marked the topic `UK Music' irrelevant
as they were only interested in the particular performers
and not in the UK Music scene in general.

\emph{Rival/foreign league teams are disliked}:
\begin{NewText}
We found for a number of cases,
that evaluators who were soccer fans,
and followed particular international soccer players,
were inferred as interested in particular soccer teams
such as Sao Paolo FC, Real Madrid, etc.
However, the evaluators marked
that they were not interested in
information from some of those particular teams.
We found that these were soccer teams 
where the players,
whom the evaluators followed, were associated.
However, it emerged that although
those players were associated with multiple teams
in their lifetime,
the evaluators were only interested in
a subset of them.
Specifically, they did not like teams
that were associated with soccer leagues
that they did not follow.
\end{NewText}

\emph{Follows celebrity but not interested}:
We found that in a number of cases,
the evaluators were inferred to be interested
in a number of Indian celebrities (Shahrukh Khan, Sachin Tendulkar, ...)
because they had followed them
and other related accounts.
However, when inquired,
the evaluators responded that although they followed certain celebrities,
they were not interested in news or information regarding them.


\subsection{Evaluating recommendation methodology}

To understand the effectiveness
of our topical recommendation strategy (Topical Reco.),
we compared it with three popular baseline recommendation algorithms.
However, as the different baseline algorithms
draw on different information sources
to generate good recommendations,
we had to create baseline specific datasets
to obtain meaningful recommendations
from the baselines.
Next we describe the baseline strategies,
and the datasets created for them.

(i) \emph{FunkSVD}: FunkSVD is a matrix factorization based
collaborative filtering algorithm~\cite{paterek-kddcup07,funksvd},
that uses regularized gradient descent SVD design.
For this work, we used the FunkSVD implementation
from the LensKit Recommendation Toolkit~\cite{ekstrand-recsys11}.
To create the user-item matrix,
to be used with the FunkSVD framework,
first we chose up to 1000 candidate users
for every target user,
as described below.
Next, the boolean user-item matrix was created
for every target user,
using the retweet and liked tweet information
of the target user as well the corresponding candidate users.
We ran FunkSVD on the user-item matrix
with 40 features and 100 training iterations per feature,
and took the top 10 recommended tweets.

To create the dataset for the FunkSVD algorithm,
we created the FunkSVD candidate user set for every target user.
For this purpose, we first crawled users who had retweeted tweets
that were liked or retweeted by the target user.
For every tweet retweeted or liked by the target user,
up to 200 such users were chosen.
Next, the crawled users were ranked
by the number of common liked/retweeted tweets
they shared with the target user.
The candidate user set was chosen
by selecting the top 1000 users from the above set.

(ii) \emph{Content Based Recommendation (Content Reco.)}:
At its base, content based recommendation algorithms
choose items to recommend,
by finding items that are most similar
to a user's already preferred items,
using some form of content similarity~\cite{lops-book11}.
To construct the content-based recommendation system,
we first rank tweets
from the set of tweets posted by a candidate set of users
based on how similar they are
to tweets posted or liked by the target user.
The top 10 tweets are then selected
for the purpose of recommendation.
We used the Jaccard similarity
for computing similarity between
the set of entities contained in the tweets.

To create the dataset for the content based recommendation method,
we first created a candidate user set for every target user,
by first crawling 1000 \emph{followees}
of every user,
the target user was following.
The \emph{followee of followee} users thus crawled,
were ranked by computing the overlap
between the user's follower set
and the target user's followee set.
The candidate user set was chosen
by selecting the top 1000 users from the above set.

(iii) \emph{Social Recommendation (Social Reco.)}:
The social recommendation process selects
the most highly retweeted or liked tweets
from the user's neighborhood~\cite{chen-chi10}.
Tweets for the social recommendation baseline
were chosen from the set of tweets retweeted or liked
by users in the candidate user set.

To create the dataset for the social recommendation algorithm,
we created target user specific candidate user sets,
by first crawling 1000 \emph{followers}
of every user,
the target user was following.
The \emph{follower of followee} users thus crawled
were ranked by computing the overlap
between the user's followee set
and the target user's followee set.
The key idea behind the above ranking being,
that users who share many followees
with the target user
are likely to have shared interests with the target user.
The candidate user set was chosen
by selecting the top 1000 users from the above set.

Additionally, for all of the above three cases,
users in the candidate sets chosen,
had following properties:
(i) they had posted a tweet within one week
from the date of evaluation,
(ii) they had more than 10 followers and 10 followees,
and (iii) they had posted at-least 10 tweets in their lifetime.
Further, candidate tweets selected for the recommendation
process were posted within one week
from the date of evaluation.

Similar to the strategy of comparing interest inference,
we presented each evaluator with four sets of top 10 tweets,
generated using the four recommendation algorithms:
our topical recommendation strategy (Topical Reco.)
and the three baselines:
FunkSVD, Content based recommendation (Content Reco.)
and Social recommendation (Social Reco.).
The evaluators were asked to rate
how much they liked each tweet
of a five point Likert scale,
ranging from `Like very much' (score 5)
to `Don't like at all' (score 1).

As with the method for comparing user's interests
we compared the results of the evaluation process
by measuring overall relevance of the methodologies
using mean average score and mean precision.
For measuring precision,
scores of 4 and 5 as were considered as relevant,
while scores between 1 and 3 were counted as irrelevant.
For measuring goodness of ranking
we used mean average precision metric and
mean normalized discounted cumulative gain (nDCG).

\begin{table}
  \begin{tabu} to \linewidth {lX[r]X[r]X[r]X[r]}

\toprule

\rowfont[c]{}

& Mean Average Score & Mean Precision & Mean Average Precision  & Mean nDCG \\

\midrule

FunkSVD       & 3.296          & 0.490          & 0.661          & 0.906          \\
Content Reco. & 3.400          & 0.532          & 0.608          & 0.901          \\
Social Reco.  & 3.307          & 0.512          & 0.622          & \textbf{0.919} \\
Topical Reco. (proposed) & \textbf{3.569} & \textbf{0.567} & \textbf{0.667} & 0.908          \\

\bottomrule
\end{tabu}
\caption{Performance comparison of baseline recommendation algorithms
(FunkSVD, content based recommendation, and social recommendation)
with the proposed Topical recommendation methodology.%
\label{tab:reco-eval}}
\end{table}



Table~\ref{tab:reco-eval} shows
the performance of the proposed Topical recommendation strategy (Topical Reco.)
when compared to the performances of the baseline strategies:
FunkSVD, Content based recommendation (Content Reco.)
and Social recommendation (Social Reco.)
Topical recommendation performs modestly better
compared to other baselines
in terms of all metrics other than mean nDCG.
In terms of mean average score and mean precision,
Topical Reco. shows 4.97\% and 6.57\% better performance
with respect to the next best algorithm Content Reco.
However, with respect to mean average precision
Topical Reco. shows only 0.91\% increase in performance
with respect to FunkSVD,
which performs second best.
Finally, we find that Topical Reco. performs 1.19\%
worse than Social recommendation,
which performs best in terms of Mean nDCG.

\begin{table}
\centering
\begin{tabu} to \linewidth {lrrl}

\toprule

\rowfont[c]{}

Feature & Coefficent & Standard Error & Signifcance \\

\midrule

(Intercept)          & 3.263  & 0.199 & *** \\
Adj. User Mean Score & 0.381  & 0.049 & *** \\
Topic Score          & 0.132  & 0.050 & ** \\
Favorite Count (log) & -0.253 & 0.078 & ** \\
Retweet Count (log)  & 0.353  & 0.087 & *** \\
Is Retweet           & -0.123 & 0.109 & \\
Has Hashtags         & 0.035  & 0.107 & \\
Has User Mentions    & -0.015 & 0.116 & \\
Has Embeded Photo    & 0.133  & 0.108 & \\
Has External Urls    & -0.014 & 0.120 & \\

\bottomrule
\end{tabu}
\caption{
Coefficents of linear regression,
for predicting the score given by the evaluators to tweets,
recommended using the Topical recommendation method.
$R^2=0.171$, $F=10.85$ ($p<10^{-14}$).
**~and *** indicate significance
at levels $\alpha=0.01$ and $\alpha=0.001$ respectively.%
\label{tab:reco-explain}}
\end{table}

To better understand the performance of our Topical recommendation algorithm,
we setup a linear regression analysis
for explaining the score given by the evaluators to tweets,
recommended by our algorithm.
The key question that we wanted to answer
from this regression analysis is that,
{\it can certain features of the recommended tweets themselves
explain the scores received by the tweets?}.
Specifically, we wanted to understand the effect of the following tweet features:
whether the tweet is a retweet
and whether the tweet contains entities such as:
hashtags, user mentions, embedded photos, and external urls.
Additionally, we included tweet based popularity measures:
number of favorites and retweets
received by the tweet at the time of evaluation.
The favorite count and retweet count of the tweets were log scaled.
Additionally, to account for the behavior of the evaluators,
we included the adjusted user mean score feature,
which is the mean of the scores given by the evaluator
to all tweets rated by him,
except the current one.
Further, we also included in the regression model,
the topic score given by the evaluator to the topic
based on which the tweet is being recommended.

Table~\ref{tab:reco-explain} shows results of the regression analysis.
We find that presence of different entities,
such as hashtags, user mentions, photos, and urls,
does {\it not} show a significant relation
with the score received by a tweet.
Also, whether a tweet is a retweet or not,
also does not seem to significantly effect the score of a tweet.
However, we find that favorite and retweet counts of tweets
have significant relations with the score received by a tweet.
Interestingly, while higher retweet count of tweets
seem to positively effect the score received by a tweet,
higher favorite counts have a negative effect
on the probability of the tweet getting a higher score.
As expected, the topic score and the evaluator's adjusted mean score
have significant effects on the score of the tweet.

\section{The \emph{What You Like} system}
\label{sec:system}

\begin{figure}
\centering
\includegraphics[width=0.7\linewidth]{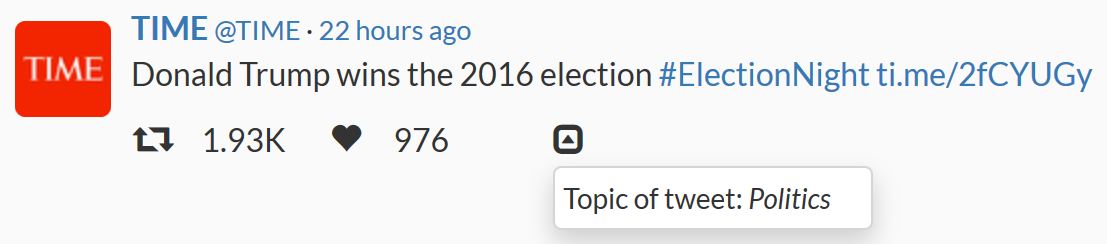}
\caption{An example of a recommendation, along with the topical explanation,
generated by the What You Like system.}
\label{fig:reco-sample}
\end{figure}

Using the recommendation methodology
presented in the current work,
we developed a web based tweet recommendation
system called \emph{What You Like},
deployed at \texttt{http://twitter-app.mpi-sws.org/what-you-like}.
The system allows any active Twitter user
to login using their Twitter account,
and discover tweets relevant to their interest.

The design of the What You Like system
consists of three distinct parts:
the tweet-topic scoring module,
the interest computation and management module,
and the recommendation and topical tweet presentation module.
We describe the modules briefly in the rest of this section.

\subsection{The tweet-topic scoring module}

The tweet scoring module is responsible for
retrieving newly posted tweets of the known topical experts
every 15 minutes,
and computing the scores of the retrieved tweets
with respect to the different topics.
The list of known topical experts is updated periodically
by crawling the Twitter Lists
created by the first 50 million Twitter users.
The crawled List data is then used
along with the latest English language Wikipedia dump,
to discover new topical experts,
using the expertise inference method
presented in Section~\ref{ssec:expertise}.
By using the efficient Topical BIM model
(described in Section~\ref{ssec:twttopic}),
for computing relevance of tweets with respect to topics,
every tweet of every expert Twitter user, posted within last 24 hours,
is scored with respect to every topic of expertise
of its author.
As of writing, the current version of the system
uses expertise information of 741 thousand experts
with expertise on 11 thousand distinct topics.

\subsection{The interest computation and management module}

When a target user uses the system for the first time,
the system fetches the list of Twitter accounts
that the target user is following,
to find out the expert followings of the user.
The expert followings of the target user is then used
for inferring her topics of interest,
by employing the interest inference method
proposed in Section~\ref{ssec:interest}.
Additionally, the system also allows the user
to directly modify her interest list,
by allowing her to remove topics from the list
and insert additional topics into it.
Over time the Twitter user may start following
additional Twitter users
and may develop new topics of interest.
To accommodate for this scenario,
the system also allows users to
manually trigger re-computation of interests
using the inference mechanism,
and incorporate into her known set of interests
any new topics discovered.

\subsection{The recommendation and topical tweet presentation module}


While the relative relevance of tweets
with respect to different topics is
computed in batch by the tweet scoring module,
the matching of user interests to tweet topics
for generating recommendation is done at run-time (i.e., when the recommendations for 
a user are to be generated).
The recommendation and topical tweet presentation module
is responsible for generating the personalized ranking
of tweets for every user,
and also allows users to view
the most relevant tweets given particular topics.
When showing a particular tweet to a user
as part of her recommendations,
the module displays the topic of interest
of the user that led to the system selecting
the tweet for the purpose of recommendation.
Figure~\ref{fig:reco-sample} shows an example tweet recommendation
generated by the system.

\section{Discussion}
\label{sec:discuss}

In this section, we discuss some observations which indicate the
practicality of the methodologies proposed in this work, and 
some potential future extensions of the present work.

\subsection{Understanding the severity in lack of rating data on Twitter}
\label{ssec:response-estimate}

State-of-the-art collaborative filtering based recommender systems
take as input the user-item rating matrix.
The effectiveness of these approaches drop
significantly in cases where sufficient rating information,
associated with most users and most items,
is not available.
Developers of recommender systems for the Twitter platform
generally use retweets and favorites/likes
as binary ratings
when recommending tweets%
~\cite{chen-sigir12,pan-recsys13}.
However, due to the well noted sparsity in retweet data,
authors often have to take alternative routes.
For example, in~\cite{chen-sigir12,pan-recsys13}
the authors chose to use user-keyword matrix,
instead of the regular user-tweet matrix
for the purposes of recommendation.

Earlier studies had noted that 44\% of Twitter users
are only passive consumers of content~\cite{Never-tweeters}.
With no rating data available
for such a large percentage of population,
the task of generating personalized recommendations
using traditional systems is very challenging.
While the above percentage shows the percentage
of users with whom no rating information is associated,
we also wanted to know the fraction of tweets
for which rating information is completely missing.

%

To estimate the percentage of tweets
that receive any rating,
that is likes or retweets,
we sampled one million original tweets
from the Twitter streaming API
for each of the seven days
during the period of June 5, 2016, to June 11, 2016.
We re-crawled those same tweets
after one month of them being tweeted.
Interestingly, we found that of the 7 million tweets
selected, only 22.7\% tweets received likes (at-least one),
while only 12.5\% tweets received retweets (at-least one).
When looking at the combined results,
we found that only 27\% of the selected tweets
have received likes or retweets.

Thus, we find the problem of data sparsity is not only
severe due to passive users,
but is even more acute when seen from the perspective
of tweets with no ratings.

\subsection{Recommending tweets not created by known topical experts}

The current study utilizes the expert sample~\cite{zafar-tweb15},
or tweets posted by known topical experts,
as the candidate set of tweets to be used for
the purposes of recommendation.
However, as shown in the third block of
Figure~\ref{fig:methodology-block-diagram},
our method can easily be used to
obtain tweet-topic scores
for a different set of candidate tweets.
The necessity of using expert tweets
is limited to the training phase
of the Topical BIM method.
Once trained however,
the Topical BIM model can then be used
to score any tweet.

The primary reason for using the expert sample
as the candidate set of tweets for this work,
is that the expert sample has been shown
to have a number of interesting properties
that make it suitable for information mining tasks%
~\cite{zafar-tweb15}.
The more interesting of these properties include,
having significantly less spam
and more informative content,
and significantly less conversational tweets.

However, many studies have been published
that have separately attacked the problems
discussed above.
This includes trying to filter out
spam and malicious content~\cite{mccord-atc11},
trying to identify information devoid of chatter~\cite{balasubramanyan-asonam13},
and trying to detect events early on Twitter~\cite{sakaki-www10}.
When used in conjunction with the above techniques,
it is possible to use Twitter streaming API,
and even the Twitter Firehose API,
as the candidate set of tweets
for doing recommendation.
In fact, when using high volume data source
such as the Twitter Firehose API,
using an efficient method for computing relevance of tweets to topics,
such as our Topical BIM methodology,
may become the only practical choice.

\subsection{Using social annotations for recommendation beyond Twitter}

The key data source utilized by this study,
that is not generally used by
traditional recommender systems,
is social annotations mined from Twitter Lists.
However, the availability of social annotations
is not limited to the Twitter microblogging platform.
A very similar feature that exists in Facebook
is known as Friend Lists, which
allows a user to create named lists of their friends on Facebook.
The Circle feature on Google+
also functions similarly.
LinkedIn allows for social annotations
in a more direct manner --
users on LinkedIn can endorse skills and
expertise of their connections~\cite{bastian-recsys14}.
Social annotations can thus be viewed
as a generally available resource,
allowing for easy extension of methods
similar to ours on other platforms.

\begin{NewText}
\subsection{Stability of relations inferred using
different recommendation methodologies}


State-of-the-art matrix factorization based
collaborative filtering algorithms
factorize the user-item matrix,
using low rank approximations,
to learn affinity of users and items
with respect to abstract latent topics~\cite{shi-csur14}.
Thus, when using such models in Twitter,
using the user-tweet matrix, one learns from the collaborative filtering model
the relationship of users and tweets with respect
to abstract topics.
However, it has been observed that most stories discussed on
Twitter change rapidly and are only discussed on Twitter
for short periods of time~\cite{asur-icwsm11}.
\citeN{asur-icwsm11} studied the trend feature on Twitter,
which capture popular evolving stories on the Twitter platform.
They noted that volumes of tweets posted,
related to different trends,
decay as a power low,
indicating short lifetimes for the stories themselves.
Additionally, the number of times
for which a story is captured in trending topic lists
also follows a power law.
The authors noted that 66\% of stories never appear
twice on the trending topic list.
Thus, it can be argued that with the ever changing
set of stories being discussed on Twitter,
abstract topics that are learned from using
collaborative filtering tools on user-tweet matrices
are also ephemeral in nature.
In other words, if models are not continuously updated
from very recent data, their performance are likely to
deteriorate quickly.

In contrast, when compared to how quickly the stories
change on Twitter, the user-user follow relationship are
much more stable. Thus, user-topic interest relationships
learned from them, such as the case in the current work,
are much more stable.

\subsection{Incorporating social annotations into MF~based
recommendation techniques}

Matrix Factorization (MF) based techniques
that create low rank approximations of data
to perform recommendation tasks
have been very popular in recent recommendation literature.
Most social recommender systems,
in addition to the user-item matrix,
incorporate user-user trust and homophily relations in their models
for performing the recommendation task%
~\cite{chen-chi10,cui-sigir11,chen-sigir12,hong-sigir12,%
zhang-www13,forsati-tois14,jiang-tkde14}.
While this paper has made different modeling choices,
it seems definitely feasible
to develop a recommendation methodology
that uses matrix factorization based modeling techniques
along with the intuitions and data sources
presented in this paper.
The key idea for such a study would be to distinguish
between experts and regular users interested in the different topics,
and use the user-expert follower matrix as opposed to the
user-user trust or homophily data.
Also, a key data source that is used in this study
is the expert-topic data, capturing the expertise information
of experts mined from social annotation data.
Incorporating the above social annotation data
would be fundamental to the success of such an approach.
\end{NewText}


\section{Conclusion}
\label{sec:conclu}


The primary contribution of this work
is in the development of a novel technique
for generating personalized tweet recommendations.
The major novelty in the proposed recommendation technique
comes from mapping both content of tweets and interests of users
to a common domain of explicit human-interpretable topics.
This is in stark contrast
to state-of-the-art model based collaborative filtering techniques,
that use abstract latent topics for the same purpose.
To execute the above task,
this paper introduces the Topical Preferential Attachment model,
that models behavior of users with respect to following of topical experts.
Based on the above model,
this paper presents an Expectation-Maximization based algorithm
to infer interests of users.
This paper also presents the Topical Binary Independence Model (Topical BIM),
that is used to efficiently compute ranked list of relevant tweets
with respect to a given topic.
The recommendation algorithm presented in this paper
combines the above two techniques to create personalized
tweet rankings for individual Twitter users.

An obvious advantage of being able to map
content of tweets and interests of users
to meaningful human-interpretable topics
is that the topics themselves can then be used
for providing simple explanations to the target user,
about why a particular item was recommended to her.
While earlier works trying to explain recommendations
have generally disassociated the process of generating explanations
from the methodology of generating recommendations%
~\cite{herlocker-cscw00},
studies have found
that in such cases users can develop
wrong intuitions about the recommendation system
which can lead to impaired user satisfaction%
~\cite{eslami-chi15}.


A major advantage of the methodology proposed
in the current work is its ability
to produce recommendations for passive Twitter users,
that is users who are passive consumers of content,
who do not tweet or retweet other's tweets.
With over 44\% of Twitter users
being only passive consumers of content~\cite{Never-tweeters},
traditional collaborative filtering and content based recommender systems,
that depend extensively on user-item rating information,
suffer tremendously due to lack of rating information
for such a large fraction of users.
Further, given the huge pace at which
tweets are produced today, and the fast pace with which
tweets lose the ability to attract attention%
~\cite{tweet-half-life},
the abstract representations of latent topics
learned by model based collaborative filtering algorithms
are likely to be extremely ephemeral in nature.
The methodology presented in the current paper
avoids the above pitfalls
by learning more stable relations of user interests
based on a user's expert followings,
which are more easily obtainable and more stable over time.

Beyond being able to recommend to passive consumers on Twitter,
this paper also demonstrated,
using a user based controlled study,
that the methodology proposed in this work
gives better results
when compared to traditional techniques,
such as matrix factorization based collaborative filtering (FunkSVD),
content based recommendation,
and social recommendation.
Evaluations from 55 different evaluators showed that
when compared using the top ten recommendations,
our methodology outperformed the baseline methods
on mean average score, mean precision,
and mean average precision (MAP).


\bibliographystyle{abbrv}
\bibliography{main}

\end{document}